\documentclass[10pt]{IEEEtran}
\usepackage{float}
\usepackage{amsmath}
\usepackage{amsthm}
\usepackage{amssymb}	
\usepackage{graphicx}
\usepackage{subfigure}
\usepackage{enumitem}
\usepackage{algorithm}
\newcommand*{\J}{\jmath}%
\usepackage{amsmath,amssymb,mathtools,bm,etoolbox}
\newtheorem{my_theorem}{Theorem}

\newtheorem{my_lemma}{Lemma}
\newtheorem{my_corollary}{Corollary}
\newtheorem{my_proposition}{Proposition}

\usepackage{color}
\usepackage{mathtools}
\usepackage{suffix}
\usepackage{breqn}
\usepackage{hhline}
\usepackage{cite}

\DeclarePairedDelimiterXPP\Aver[1]{\mathbb{E}}{[}{]}{}{
	
	#1
}
\DeclarePairedDelimiterX\MeijerM[3]{\lparen}{\rparen}%
{\,#3\delimsize\vert\begin{matrix}#1 \\ #2\end{matrix}}
\newcommand\MeijerG[8][]{%
	G^{\,#2,#3}_{#4,#5}\MeijerM[#1]{#6}{#7}{#8}}
\newcommand{\diff}{\mathop{}\!d}

\WithSuffix\newcommand\MeijerG*[7]{%
	G^{\,#1,#2}_{#3,#4}\MeijerM*{#5}{#6}{#7}}

\title{Multiple RIS-Assisted  Mixed FSO-RF  Transmission Over Generalized Fading  Channels}
\author{Vinay Kumar Chapala,~\IEEEmembership{Graduate Student Member,~IEEE} and  S.~M.~Zafaruddin,~\IEEEmembership{Senior Member,~IEEE} 
		\thanks{ This work was supported in part by the Science and Engineering Research Board (SERB), Government of India, under MATRICS Grant MTR/2021/000890.}	
	\thanks{A conference version of the paper analyzing the  RF link consisting of  multiple RIS over dGG fading was presented at the 2022 IEEE 95th Vehicular Technology Conference (VTC 2022-Spring), Helsinki, Finland, 19-22 June 2022  \cite{chapala2022vtcmultiris}. }
	\thanks{The authors are  with  the Department of Electrical and Electronics Engineering, Birla Institute of Technology and Science, Pilani, Pilani-333031, Rajasthan, India, Email: p20200110@pilani.bits-pilani.ac.in, syed.zafaruddin@pilani.bits-pilani.ac.in.}
}

\begin{document}
	\maketitle 
	\begin{abstract}
  In this paper, we analyze the performance of a  reconfigurable intelligent surface (RIS)-assisted multi-hop transmission by employing multiple  RIS units to enable favorable communication for a  mixed free-space optical (FSO) and radio-frequency (RF) system. We consider a single-element RIS since it is hard to realize phase compensation for multiple-element RIS in the multi-hop scenario. We develop statistical results for the product of the signal-to-noise ratio (SNR) of the cascaded multiple RIS-equipped wireless communication. We use decode-and-forward (DF)  and fixed-gain (FG) relaying protocols to mix multi-RIS transmissions over RF and FSO technologies and derive probability density and distribution functions for both the relaying schemes by considering independent and non-identical double generalized gamma (dGG)  distribution models for RF transmissions   with line-of-sight (LOS) and inverse-Gamma shadowing effect and atmospheric turbulence for FSO system combined with pointing errors. We analyze the outage probability, and average bit-error rate (BER) performance of the considered system. We also present an asymptotic analysis of the outage probability using gamma functions to provide insight into the considered system in the high SNR regime. We use computer simulations to validate the derived analytical expressions and demonstrate the performance for different system parameters on the RIS-assisted multi-hop transmissions for a vehicular communication system.
	\end{abstract}				

\section{Introduction}
Traditionally, wireless communication systems are optimized based on the principle that channel environments cannot
be configured. Adaptive modulation, optimization of transmit power, diversity techniques, and
cooperative relaying are methods to mitigate channel impairments such as  channel fading, path loss, and shadowing effect. It is intriguing to accept that signal propagation in an unknown environment due to
scattering, reflection, and refraction of electromagnetic waves can be controlled. Reconfigurable intelligent surface (RIS) is evolving as a next-generation technology for reliable wireless and vehicular communications  \cite{Renzo2019,Basar2019_access,Qingqing2021,Yishi_2021}. In general, RIS modules consist  planar structures of metasurfaces that can be intelligently  programmed to steer the incident waves in a particular direction. An  RIS unit accommodating many meta elements to reflect the signal is being considered a potential alternative to cooperative relaying for 6G wireless systems. In recent years, free-space optical (FSO) for  backhaul/fronthaul and radio frequency (RF) for broadband access have been considered  as a potential architecture for next generation wireless communications \cite{Trichili20}.  The FSO is potential technology with a higher contiguous unlicensed optical spectrum, which can be employed for secured high data rate transmissions for backhaul applications. The RIS can help FSO and RF links to maintain line-of-sight (LOS) connectivity for an improved performance for both vehicular and terrestrial  communications.  

 Recent research focus on  the applicability of a single RIS (mostly with multiple elements) based wireless systems for intelligent vehicular communications \cite{Wang_2020_outage,Kehinde_2020}, low-frequency radio-frequency (RF) \cite{trigui2020_fox, du2021}, optical wireless communications (OWC) \cite{Jamali2021, Najafi2019, ndjiongue2021,chapala2021unified}, and high-frequency terahertz (THz) wireless systems \cite{du2020_thz, Dovelos2021, Chapala2021THz}. 
 However, the use  of multiple RIS module can provide  reliable connectivity for different situations.  Ozcan {\emph {et al.}}  in \cite{Ozcan2021} provided optimized algorithms for  placing multiple RISs alongside the road for vehicle-to everything (V2X) infrastructure.   Huang {\emph {et al.}} \cite{huang2021} suggested  hybrid beamforming  under  multi-hop  scenario to extend the transmission range at terahertz frequencies.  Boulogeorgos {\emph {et al.}} \cite{boulogeorgos2021cascaded} analyzed the performance of a cascaded  FSO system assisted by multiple  RIS modules (with single-element) considering Gamma-Gamma atmospheric turbulence and  pointing errors. They applied the method of induction  to derive the probability  density function (PDF) and cumulative distribution function (CDF). It is known that the double generalized Gamma (dGG) distribution  for the atmospheric turbulence in FSO system  is superior to the Gamma-Gamma model since the dGG model is valid under all range of turbulence conditions (weak to strong) for both plane and spherical waves propagation \cite{Kashani2015,  AlQuwaiee2015,  Ashrafzadeh2020, RahmanTVT2021}.  Further, the dGG model is an appropriate model for   radio propagation  for V2V communications over RF frequencies \cite{Petros2018}.  The dGG is a generalized model consisting most of the existing statistical models as particular cases.   To this end, it should be mentioned that the induction method of \cite{boulogeorgos2021cascaded} may not be applicable for the cascaded dGG fading model to conduct performance analysis for wireless transmission empowered by multiple  RIS units.

 There has been some study on the performance of a single RIS (equipped with multiple elements) assisted mixed FSO-RF systems \cite{Yang_2020_fso,Sikri21, Salhab_2021, Liang2020_vlc}. In \cite{Yang_2020_fso,Sikri21,Salhab_2021}, the authors used the near-optimal decode-and-forward (DF) relaying protocol to integrate an FSO  transmission link and  RIS-equipped RF link distributed according to the simpler Rayleigh fading model.  In \cite{Salhab_2021}, authors analyzed the performance of a RIS-equipped source mixed RF/FSO system by considering RIS located close to the source in the RF link. It should be mentioned that fixed-gain amplify-and-forward (AF) relaying is well acknowledged for its desirable characteristics of lower computational complexity and the fact that it does not require continuous channel monitoring for decoding at the relay \cite{Hasna_2004_AF}.  Further, deriving  analytical results for fixed-gain AF relaying with general channel models is  challenging compared with the DF protocol. The authors in \cite{Liang2020_vlc} investigated the outage probability  performance of a fixed-gain relayed system for the mixed RF-visible light communication   for an indoor environment.   To the best of authors' knowledge, a multiple RIS empowered mixed FSO-RF relayed system over dGG fading channels has not been reported in the literature.

This paper analyzes  the performance of a mixed FSO-RF system by employing multiple  RISs in both the links realizing multi-hop transmissions for reliable connectivity.  We assume the dGG turbulence model and zero-boresight pointing errors for the FSO link and the dGG distribution to model the signal fading combined with the effect of shadowing distributed as inverse-Gamma (IG) for vehicular transmissions over RF along with the LOS link and a statistical model for mobility. The major contributions of the paper are listed as follows:
 \begin{itemize}[leftmargin=*]
 	\item  We consider a single-element RIS since  it is hard for phase compensation to be realized for multiple-element RIS in a multi-hop scenario. 
 	\item  We develop statistical results for the product of the signal-to-noise ratio (SNR) of the cascaded multiple RIS-equipped wireless communication.
 \item  We derive PDF and CDF of the fixed-gain (FG) relaying using Fox's H-function  to mix multi-RIS empowered   vehicular transmissions over RF  and FSO backhaul with zero-boresight pointing errors  considering independent and non-identical (i.ni.d) double generalized gamma (dGG) distribution function without any constraint on its fading parameters.   
  \item  We analyze the performance of the AF-assisted system by deriving exact analytical expressions of the	outage probability and average bit-error-rate (BER) in terms of trivariate Fox's H-function.   We also analyze the performance   by integrating the FSO and RF systems using the DF relaying in terms of bivariate Fox's H-function. 
   \item  We present asymptotic analysis of the outage probability in terms of simpler Gamma functions to develop  better insight on the system performance  in the high SNR regime. 
   \end{itemize}

\begin{figure*}[tp]
	\centering
		\vspace{-4.5cm}
	{\includegraphics[scale=0.4]{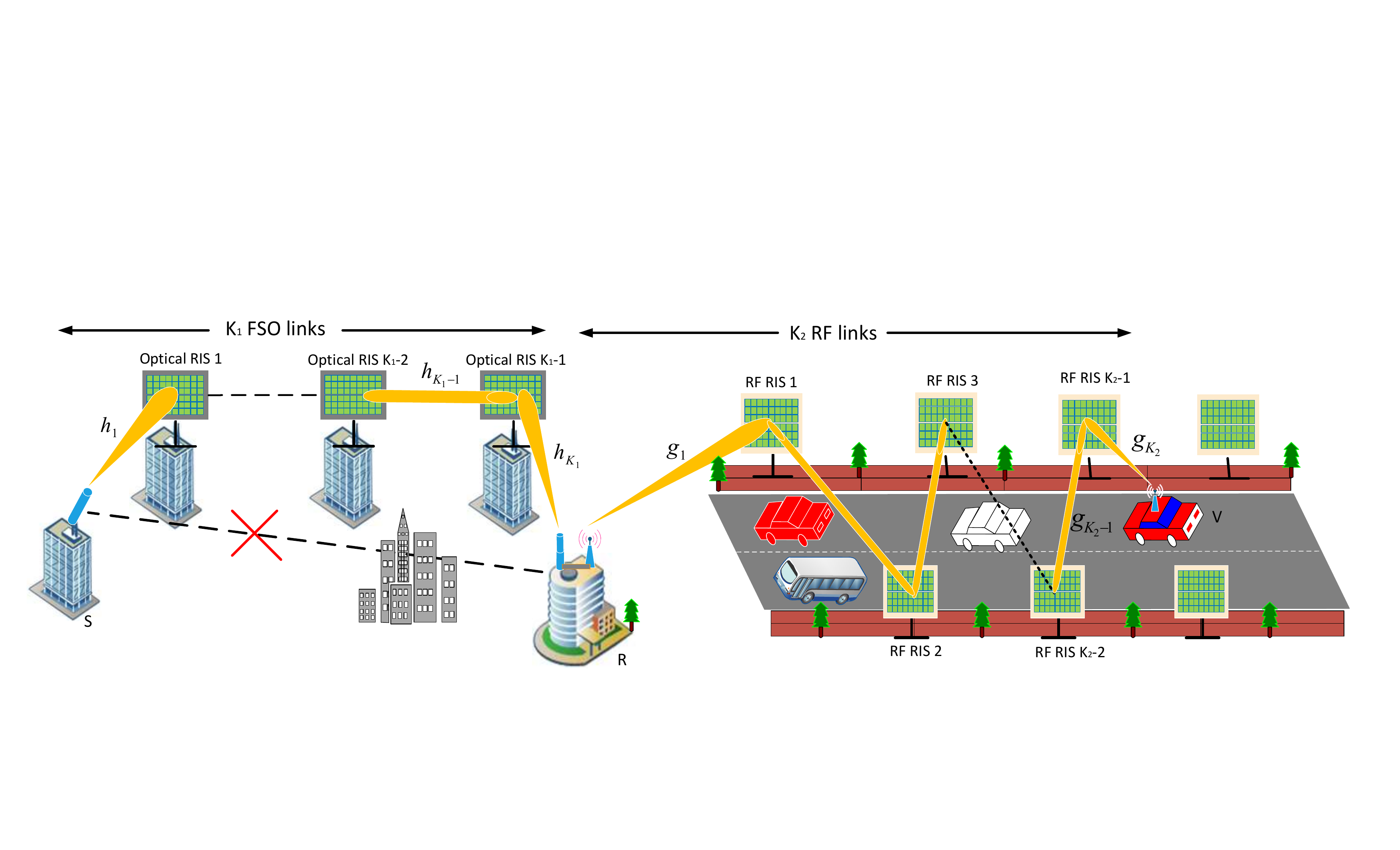}}
		\vspace{-2.0cm}
		\caption{A block diagram description for multiple RIS empowered vehicular communications.}
	\label{fig:system_model}
\end{figure*}

\section{Description of the System Model}
	Consider a multi-hop mixed FSO/RF system  assisted by multiple RISs in both FSO and RF links, as depicted in Fig. \ref{fig:system_model}. The source ($S$) desires to connect with a moving vehicle ($V$). We consider a communication link comprising of backhaul/fronthaul on the FSO technology and vehicular link on the RF. A relay ($R$) with the functionality of AF/DF  is deployed to integrate the FSO and RF technologies.   The relay contains a photodetector for receiving FSO signals and a frequency down-converter for low frequency RF communications. We employ multiple  optical RIS to assist the communication from the source to the relay over a building-to-building backhaul link since there is no direct link between them for the line-of-sight FSO transmissions.  In FSO links, the use of multiple RIS modules may reduce the pointing error in each hop due to near-perfect beam alignment with subsequent use of multiple RIS. Thus, the source  communicates data to the  relay through $K_{1}$ hops using $K_{1}-1$ optical RIS modules. Further,  we employ $K_2-1$ RF RIS modules perpendicular to the road such that the desired vehicle is connected to any one of the RIS modules.  With the multiple RF-RIS deployment, the effect of deep shadowing can be reduced due to the close proximity of the moving vehicle to RIS.
	
	We assume that each optical RIS  reflects the incident beam towards subsequent optical RIS until the last RIS directs the beam towards the relay. We assume perfect knowledge of the channel at each RIS to perform phase control. Indeed, practical algorithms are needed to estimate the channel at each RIS to better reflect the incident signal in a desired direction. In a recent paper \cite{do2021multirisaided}, a different system configuration has been considered by employing  multiple RIS such that the destination directly receives signals from each RIS independent of other RIS. We consider single-element RIS in the multi-hop situation as phase compensation in a multiple-element RIS under a multihop deployment scenario is hard to realize. Thus, the  signal $y_{R}$ at the relay unit through $(K_{1}-1)$ optical RISs is given by
\begin{equation}
\label{eq:main1_zafar}
	y_{R} = \prod_{i=1}^{K_{1}} h_{i} A^{FSO}_ie^{j\theta^{FSO}_i} s + w_{R} 
\end{equation}
where $h_{1}$ denotes the channel coefficient from the transmission source to the first-optical RIS whereas  $h_{i}$ is the channel coefficient from $(i-1)$-th optical RIS to $i$-optical RIS. Here,     $h_{K_{1}}$ is the channel coefficient of the last $(K_{1}-1)$-optical RIS to the relay.  $s$ is the transmitted signal  with power $P$,  and $w_{R}$ is the additive white Gaussian noise (AWGN) with variance $\sigma^2_R$.   The terms $A_i$ and $\theta^{FSO}_i$ denote the gain and phase of the $i$-th FSO-RIS module, respectively.

Similarly, the signal at destination vehicle connected through $(K_{2}-1)$-th RIS along with LOS signal is given by
\begin{equation}\label{eq:rf_main_zaf}
	y_{} = \prod_{i=1}^{K_{2}} g_{i}A^{RF}_ie^{j\theta^{RF}_i} s + g_{\rm LOS} s + w_{V} 
\end{equation}
where $w_{V}$ is the AWGN at the destination with noise variance $\sigma^2_V$, $g_{1}$ is the channel coefficient from the relay to the first RIS, $g_{i}$ is the channel from $(i-1)$-th RIS to $i$-RIS, and $g_{K_{2}}$ is the channel coefficient of the last $(K_{2}-1)$-RIS to destination vehicle. The terms $A_i^{RF}$ and $\theta^{RF}_i$ denote the gain and phase of the $i$-th RF-RIS module, respectively. Here, $g_{\rm LOS}$ is the channel coefficient of LOS link from the relay to destination vehicle.

We consider the dGG distribution to model the atmospheric turbulence for the FSO  and as multi-path fading model for the RF link. As such, the dGG is the product of two generalized Gamma functions. The PDF of the  generalized Gamma function is given as
\begin{equation}\label{eq:pdf_gen_gamma}
	f_{\chi}(x) = \frac{\alpha_{}x^{\alpha_{}\beta_{}-1}}{(\frac{\Omega_{}}{\beta_{}})^{\beta_{}}\Gamma(\beta_{})} \exp\big(-\frac{\beta_{}}{\Omega_{}}x^{\alpha_{}}\big) 
\end{equation}
where $\alpha$, $\beta$ are shaping parameters of the Gamma distribution  and $\Omega=\big(\frac{\mathbb{E}[\chi^{2}]\Gamma(\beta)}{\Gamma(\beta+2/\alpha)}\big)^{\alpha/2}$ is the $\alpha$-root mean value parameter.

In addition to the atmospheric turbulence, we include pointing errors in each hop of the FSO link such that the resultant channel becomes $h_{i} = h_{i}^{(l)} h_{i}^{(t)} h_{i}^{(p)}$, where subscripts $(l)$, $(t)$ and $(p)$ denote path gain, atmospheric turbulence coefficient and pointing error coefficient. 
To characterize  pointing errors statistically, we use the recent model proposed in \cite{Wang2020}:
\begin{equation}
	\begin{aligned}
		f_{h_{i}^{(p)}}(x) &= \frac{\rho_{i}^2}{A_{0,i}^{\rho_{i}^2}}x^{\rho_{i}^{2}-1},0 \leq x \leq A_{0,i},
	\end{aligned}
	\label{eq:pointing_error_pdf}
\end{equation}
where the term $A_{0,i}$ denotes the fraction of collected power at the receiver aperture and $\rho_{i}^2 = {\frac{\omega^2_{z_{\rm eq}}}{\xi}}$ where $\omega_{z_{\rm eq}}$ is the equivalent beam-width at the receiver.  The lower value of $\rho_{i}$ and $A_{0,i}$ indicates higher pointing error. An extension to the non-zero boresight pointing errors is also possible, however,  with more  complicated analytical derivations \cite{Jung2020GPE}. 

Similarly, we denote the channel coefficient between the $(i-1)$-th and $i$-th RIS as  $g_{i} = g_{i}^{(l)} g_{i}^{(s)} g_{i}^{(f)}$, where subscripts $(l)$, $(s)$ and $(f)$ denote path loss, shadowing and short-term fading coefficients of RF links. Moreover, we include the analysis   of direct link  represented as $g_{\rm LOS} = g_{\rm LOS}^{(l)} g_{\rm LOS}^{(s)} g_{\rm LOS}^{(f)}$. Note that many  models exist in the literature to characterize the effect of shadowing. However, we use more recently proposed IG model which is validated through practical measurements and analytical tractability. We consider the IG distribution to model the effect of shadowing whose PDF is obtained using \cite[eq. 5]{Rami2021IGG} and through transformation of random variables, we get
\begin{equation}
	\begin{aligned}
		f_{g_{i}^{(s)}}(x) &= \frac{2(m_{i}-1)}{\Gamma(m_{i})}x^{2m_{i}-1} \exp\big(-\frac{m_{i}-1}{x^{2}}\big)
	\end{aligned}
	\label{eq:ig_pdf}
\end{equation}

The path loss component of each RF link is considered as $g_{i}^{(l)}= (\frac{c}{4\pi f_{c}d_{K_{2}}})^{}$, where $d_{K_{2}}$ is the length of each hop, $c=3 \times 10^8$ \mbox{m/s}, and $f_{c}$ is the center frequency. Note that the path-loss $g_{K_{2}}^{(l)}$ between the RIS and destination vehicle is random considering the movement of the vehicle. We denote that path loss component of last hop as $g_{K_{2}}^{(l)}=r^{-\frac{a}{2}}$, where $a, 2\le a\le 5$ is the path loss exponent factor. We assume random way point (RWP) mobility model for the path loss with probabilistic distance $r$ of the last hop, $0\le r \le d_{K_{2}}$ \cite{Govindan_mobility}:
	\begin{eqnarray}\label{eq:mobility_power_pdf}
		f_{r}(x) = 6 \frac{x^{}}{d_{K_{2}}^{2}} -6 \frac{x^{2}}{d_{K_{2}}^{3}}
	\end{eqnarray}

It should be mentioned that the received signal in \eqref{eq:main1_zafar} and \eqref{eq:rf_main_zaf} depends on the phase error $\theta$, the SNR (the metric taken in the manuscript) becomes independent of phase error.   Thus,  there is no effect of phase on the SNR with single-element RIS as the effective channel is product of channel coefficients in each hop.
\section{Statistical Results for Cascaded Channels}
In this section, we develop statistical results of the multihop channels for both FSO and RF to facilitate average-value performance analysis of different metrics such as average BER.
First, we require PDF and CDF of the cascaded FSO channels $h=\prod_{i=1}^{K_{1}}h_{i}^{(tp)} = \prod_{i=1}^{K_{1}} h_{i}^{(t)} h_{i}^{(p)}$ and cascaded RF channels $g=\prod_{i=1}^{K_{2}} g_{i}^{(f)}$. 

In the following two propositions, we express the PDF of  dGG $g_{i}^{(f)}$ and product of the dGG with pointing errors $h_{i}^{(tp)}$ in terms of Fox's-H function without the limitation of integer-valued fading parameter due to the Meijer-G representation, as presented earlier \cite{Kashani2015,AlQuwaiee2015}. 
\begin{my_proposition}
	The PDF of dGG distributed channel $h_{i}^{(t)}=\chi_{i,1}\chi_{i,2}$, where $\chi_{i,1}\thicksim \mathcal{GG}(\alpha_{i,1},\beta_{i,1},\Omega_{i,1})$ and $\chi_{i,2}\thicksim \mathcal{GG}(\alpha_{i,2},\beta_{i,2},\Omega_{i,2})$ can be expressed as
	\begin{align}\label{eq:dgg_pdf}
	   &f_{h_{i}^{(t)}}(x) = \frac{x^{\alpha_{i,2}\beta_{i,2}-1}}{(\frac{\Omega_{i,1}}{\beta_{i,1}})^{\frac{\alpha_{i,2}\beta_{i,2}}{\alpha_{i,1}}}(\frac{\Omega_{i,2}}{\beta_{i,2}})^{\beta_{i,2}}\Gamma(\beta_{i,1})\Gamma(\beta_{i,2})} \nonumber \\ &H_{0,2}^{2,0} \bigg[\begin{array}{c} \zeta_{i} x \end{array} \big\vert \begin{array}{c} - \\ (0,\frac{1}{\alpha_{i,2}}),(\frac{\alpha_{i,1}\beta_{i,1}-\alpha_{i,2}\beta_{i,2}}{\alpha_{i,1}},\frac{1}{\alpha_{i,1}})\end{array}\bigg]
	\end{align}	
	where $\zeta_{i} = \big(\frac{\beta_{i,2}}{\Omega_{i,2}}\big)^{\frac{1}{\alpha_{i,2}}} \big(\frac{\beta_{i,1}}{\Omega_{i,1}}\big)^{\frac{1}{\alpha_{i,1}}}$. 

\end{my_proposition}
\begin{IEEEproof}We use the  Mellin's convolution to get the PDF of the product of two random variables as	$f_{h_{i}^{(t)}}(x) = 	\int_{0}^{\infty} \frac{1}{u} f_{\chi_{i,1}}(\frac{x}{u}) f_{\chi_{i,2}}(u) \diff u$, where the limits of the integral are selected  using the inequalities $0\leq \frac{x}{u}\leq \infty$ and $0\leq u\leq \infty$ since  $\chi_{1}\in [0,\infty)$ and $\chi_{2}\in [0,\infty)$:
	\begin{align}\label{eq:dgg_pdf_1}
		&f_{h_{i}^{(t)}}(x)= \frac{\alpha_{i,1}\alpha_{i,2}x^{\alpha_{i,1}\beta_{i,1}-1}}{(\frac{\Omega_{i,1}}{\beta_{i,1}})^{\beta_{i,1}}(\frac{\Omega_{i,2}}{\beta_{i,2}})^{\beta_{i,2}}\Gamma(\beta_{i,1})\Gamma(\beta_{i,2})}  \int_{0}^{\infty} u^{\alpha_{i,2}\beta_{i,2}-1}\nonumber \\ & u^{-\alpha_{i,1}\beta_{i,1}} \exp\big(-\frac{\beta_{i,1}x^{\alpha_{i,1}}}{u^{\alpha_{i,1}}\Omega_{i,1}} \big) \exp\big(-\frac{\beta_{i,2}u^{\alpha_{i,2}}}{\Omega_{i,2}}\big) \diff u
	\end{align}
	Using $u^{-\alpha_{i,1}}=t$ and  the Meijer-G equivalent of the  exponential function, we get
	\begin{align}\label{eq:dgg_pdf_2}
		&f_{h_{i}^{(t)}}(x) = \frac{\alpha_{i,2}x^{\alpha_{i,1}\beta_{i,1}-1}}{(\frac{\Omega_{i,1}}{\beta_{i,1}})^{\beta_{i,1}}(\frac{\Omega_{i,2}}{\beta_{i,2}})^{\beta_{i,2}}\Gamma(\beta_{i,1})\Gamma(\beta_{i,2})}  \int_{0}^{\infty} t^{\frac{-\alpha_{i,2}\beta_{i,2}}{\alpha_{i,1}}}\nonumber \\    &\hspace{0mm}t^{\beta_{i,1}-1}G_{0,1}^{1,0}\bigg[\begin{array}{c} \frac{\beta_{i,1}x^{\alpha_{i,1}}}{\Omega_{i,1}} t\end{array} \big\vert \begin{array}{c} - \\ 0\end{array}\bigg]  G_{1,0}^{0,1}\bigg[\begin{array}{c} \frac{\Omega_{i,2}}{\beta_{i,2}}t^{\frac{\alpha_{i,2}}{\alpha_{i,1}}}\end{array} \big\vert \begin{array}{c} 1 \\ -\end{array}\bigg] \diff t \hspace{-2mm}
	\end{align}
	\normalsize
We apply the identity \cite[07.34.21.0012.01]{Meijers} in \eqref{eq:dgg_pdf_2} to get \eqref{eq:dgg_pdf}. The PDF of dGG distributed RF channel $g_{i}^{(f)}=\chi_{i,3}\chi_{i,4}$, where $\chi_{i,3}\thicksim \mathcal{GG}(\alpha_{i,3},\beta_{i,3},\Omega_{i,3})$ and $\chi_{i,4}\thicksim \mathcal{GG}(\alpha_{i,4},\beta_{i,4},\Omega_{i,4})$ has also of the form given by \eqref{eq:dgg_pdf}.
\end{IEEEproof}
Unlike the Meijer's-G representation in the original work in \cite{Kashani2015}, the derived PDF in \eqref{eq:dgg_pdf} does not have integer-valued constraint  on  fading parameters of the dGG. The use of Fox's-H function for exact and asymptotic  expressions for generalized fading models for wireless links is a  popular choice among researchers.  Recently, the computational software MATHEMATICA  introduced the function \emph{FoxH}  for the computation of Fox's-H function.

As such, dGG model becomes Rayleigh fading with $\alpha_{i,1}=\alpha_{i,2}=1$ and $\beta_{i,1}=\beta_{i,2}=2$ and represents Nakagami-m fading for $\alpha_{i,1}=\alpha_{i,2}=1$ and $\beta_{i,1}=\beta_{i,2}=2m$.

Next, we derive the PDF of the FSO link with the combined effect of dGG atmospheric turbulence and pointing errors:
\begin{my_proposition}
	If the pointing error parameter $h_{i}^{(p)}$ is distributed according to \eqref{eq:pointing_error_pdf}, then the PDF of the single FSO link  $h_{i}^{(tp)}=h_{i}^{(t)}h_{i}^{(p)}    $  with atmospheric turbulence distributed as dGG and pointing errors is represented as
	\begin{align}\label{eq:dgg_pointing_error_pdf}
		&f_{h_{i}^{(tp)}}(x) =  \frac{\rho_{i}^2x^{\alpha_{i,2}\beta_{i,2}-1}}{A_{0,i}^{\alpha_{i,1}\beta_{i,2}}(\frac{\Omega_{i,1}}{\beta_{i,1}})^{\frac{\alpha_{i,2}\beta_{i,2}}{\alpha_{i,1}}}(\frac{\Omega_{i,2}}{\beta_{i,2}})^{\beta_{i,2}}\Gamma(\beta_{i,1})\Gamma(\beta_{i,2})} \nonumber \\ &H_{1,3}^{3,0} \big[\begin{array}{c}  \frac{\zeta_{i}x}{A_{0,i}}\end{array} \big\vert \begin{array}{c} (\rho_{i}^{2}-\alpha_{i,2}\beta_{i,2}+1,1) \\ V_{1} \end{array}\big]\hspace{-2mm}
	\end{align}	
	\normalsize
	where $\zeta_{i} = \big(\frac{\beta_{i,2}}{\Omega_{i,2}}\big)^{\frac{1}{\alpha_{i,2}}} \big(\frac{\beta_{i,1}}{\Omega_{i,1}}\big)^{\frac{1}{\alpha_{i,1}}}$ and $V_{1}=(0,\frac{1}{\alpha_{i,2}}),(\beta_{i,1}-\frac{\alpha_{i,2}\beta_{i,2}}{\alpha_{i,1}},\frac{1}{\alpha_{i,1}}),(\rho_{i}^{2}-\alpha_{i,2}\beta_{i,2},1)$.
\end{my_proposition}
\begin{IEEEproof}
	The combined PDF of dGG and pointing error can be expressed using the standard result on the product of two random variables
	\begin{equation}\label{eq:dgg_pointing_error_pdf_1}
		f_{h_{i}^{(tp)}}(x) = \int_{0}^{A_{0,i}} \frac{1}{u} f_{h_{i}^{(t)}}(\frac{x}{u})f_{h_{i}^{(p)}}(u) du
	\end{equation}
	where the integral limits are based on the inequalities $0\leq \frac{x}{u}\leq \infty$ and $0\leq u\leq A_{0}$ since  $h_{i}^{(t)}\in [0,\infty)$ and $h_{i}^{(p)}\in [0,A_{0}]$.
	Substituting  \eqref{eq:dgg_pdf} and \eqref{eq:pointing_error_pdf} in \eqref{eq:dgg_pointing_error_pdf_1} along with the definition of Fox-H function and applying the Fubini's theorem to interchange the integrals resulting into
	\begin{align}\label{eq:dgg_pointing_error_pdf_2}
		&	f_{h_{i}^{(tp)}}(x) = \frac{\alpha_{i,2}x^{\alpha_{i,2}\beta_{i,2}-1}}{(\frac{\Omega_{i,1}}{\beta_{i,1}})^{\frac{\alpha_{i,2}\beta_{i,2}}{\alpha_{i,1}}}(\frac{\Omega_{i,2}}{\beta_{i,2}})^{\beta_{i,2}}\Gamma(\beta_{i,1})\Gamma(\beta_{i,2})} \frac{\rho_{i}^2}{A_{0,i}^{\rho_{i}^2}}   \nonumber \\&  \frac{1}{2\pi\J} \int_{\mathcal{L}} \bigg(\frac{\beta_{i,2}}{\Omega_{i,2}} \big(\frac{\beta_{i,1}}{\Omega_{i,1}}\big)^{\frac{\alpha_{i,2}}{\alpha_{i,1}}} (x)^{\alpha_{i,2}}\bigg)^{s} \Gamma(\beta_{i,1}-\frac{\alpha_{i,2}\beta_{i,2}}{\alpha_{i,1}}-\frac{\alpha_{i,2}}{\alpha_{i,1}}s) \nonumber \\ & \Gamma(-s)\bigg(\int_{0}^{A_{0,i}} u^{\rho_{i}^{2}-\alpha_{i,2}\beta_{i,2}-\alpha_{i,2}s-1} du\bigg) \diff s 
	\end{align}
	\normalsize
	The inner integral of \eqref{eq:dgg_pointing_error_pdf_2} is solved as $\int_{0}^{A_{0,i}} u^{\rho_{i}^{2}-\alpha_{i,2}\beta_{i,2}-\alpha_{i,2}s-1} du$ = $\frac{A_{0,i}^{\rho_{i}^{2}-\alpha_{i,2}\beta_{i,2}-\alpha_{i,2}s}}{\rho_{i}^{2}-\alpha_{i,2}\beta_{i,2}-\alpha_{i,2}s}$ = $A_{0,i}^{\rho_{i}^{2}-\alpha_{i,2}\beta_{i,2}-\alpha_{i,2}s} \frac{\Gamma(\rho_{i}^{2}-\alpha_{i,2}\beta_{i,2}-\alpha_{i,2}s)}{\Gamma(\rho_{i}^{2}-\alpha_{i,2}\beta_{i,2}-\alpha_{i,2}s+1)}$. Further, we substitute the inner integral in  \eqref{eq:dgg_pointing_error_pdf_2}, and use the integral representation of the  Fox's H function  and  the identity  \cite[{eq. }1.59]{M-Foxh} to  get  \eqref{eq:dgg_pointing_error_pdf}.
\end{IEEEproof}
As a sanity check, we can integrate $\int_{0}^{\infty} f_{h_{i}^{(tp)}}(x) dx$ to validate the derived PDF in \eqref{eq:dgg_pointing_error_pdf} as
\begin{align}\label{eq:dgg_pointing_pdf_proof}
	&\hspace{0mm}=\frac{\rho_{i}^{2}\Gamma(\beta_{i,2}) \Gamma(\beta_{i,1})(\frac{\beta_{i,2}}{\Omega_{i,2}})^{-\beta_{i,2}} (\frac{\beta_{i,1}}{\Omega_{i,1}})^{\frac{-\alpha_{i,2}\beta_{i,2}}{\alpha_{i,1}}}}{(\frac{\Omega_{i,1}}{\beta_{i,1}})^{\frac{\alpha_{i,2}\beta_{i,2}}{\alpha_{i,1}}}(\frac{\Omega_{i,2}}{\beta_{i,2}})^{\beta_{i,2}}\Gamma(\beta_{i,1})\Gamma(\beta_{i,2})\rho_{i}^{2}}  = 1
\end{align}
\normalsize
Note that PDF of  \cite{AlQuwaiee2015} is a  particular case of the Fox-H representation of the  derived PDF in \eqref{eq:dgg_pointing_error_pdf}.

We then derive the PDF of the combined effect of IG shadowing and dGG fading for the RF link:
	\begin{my_proposition}
		The PDF of the single RF link  $g_{i}^{(sf)}=g_{i}^{(s)}g_{i}^{(f)} $  with the combined effect of shadowing and dGG is given by
		\begin{align}\label{eq:dgg_shadowng_pdf}
			&f_{g_{i}^{(sf)}}(x) =  \frac{x^{-1}(m_{i}-1)^{m_{i}}}{\Gamma(\beta_{i,3})\Gamma(\beta_{i,4})} \nonumber \\ &\hspace{-2mm}H_{1,2}^{2,1} \big[\begin{array}{c}  \frac{\zeta_{i}}{\sqrt{(m_{i}-1)^{}}}x\end{array} \big\vert \begin{array}{c} (1+m_{i},\frac{1}{2}) \\ (\beta_{i,4},\frac{1}{\alpha_{i,4}}),(\beta_{i,3},\frac{1}{\alpha_{i,3}})\end{array}\big]\hspace{-2mm}
		\end{align}
		\normalsize
		where $\zeta_{i} = \big(\frac{\beta_{i,3}}{\Omega_{i,3}}\big)^{\frac{1}{\alpha_{i,3}}} \big(\frac{\beta_{i,4}}{\Omega_{i,4}}\big)^{\frac{1}{\alpha_{i,4}}}$.
	\end{my_proposition}
	\begin{IEEEproof}
		The combined PDF of dGG and shadowing can be expressed as
		\begin{equation}\label{eq:dgg_shadowng_pdf_1}
			f_{g_{i}^{(sf)}}(x) = \int_{0}^{\infty} \frac{1}{u} f_{g_{i}^{(f)}}(\frac{x}{u})f_{g_{i}^{(s)}}(u) du
		\end{equation}
		Substituting  \eqref{eq:dgg_pdf} and \eqref{eq:ig_pdf} in \eqref{eq:dgg_shadowng_pdf_1} with the definition of Fox-H function and interchanging the integrals as per Fubinis theorem to get
		\begin{align}\label{eq:dgg_shadowng_pdf_2}
			&	f_{g_{i}^{(sf)}}(x) = \frac{x^{\alpha_{i,4}\beta_{i,4}-1}}{(\frac{\Omega_{i,3}}{\beta_{i,3}})^{\frac{\alpha_{i,4}\beta_{i,4}}{\alpha_{i,3}}}(\frac{\Omega_{i,4}}{\beta_{i,4}})^{\beta_{i,4}}\Gamma(\beta_{i,3})\Gamma(\beta_{i,3})}    \nonumber \\&\frac{1}{2\pi\J} \int_{\mathcal{L}}    \Gamma(-\frac{s}{\alpha_{i,4}})\Gamma(\frac{\alpha_{i,3}\beta_{i,3}-\alpha_{i,4}\beta_{i,4}}{\alpha_{i,3}}-\frac{s}{\alpha_{i,3}})  \big(\zeta_{i} x^{}\big)^{s}\nonumber \\ &\bigg(\int_{0}^{\infty} u^{2m_{i}-\alpha_{i,4}\beta_{i,4}-s-1} e^{-(m_{i}-1)u^{-2}} du\bigg) \diff s 
		\end{align}
		\normalsize
		The inner integral of \eqref{eq:dgg_shadowng_pdf_2} is solved using the substitution $t=(m_{i}-1)u^{-2}$ as $(m_{i}-1)^{\frac{2m_{i}-\alpha_{i,4}\beta_{i,4}-s}{2}}\int_{0}^{\infty} t^{\frac{s+\alpha_{i,4}\beta_{i,4}-2m_{i}}{2}-1} e^{-t}dt$ = $(m_{i}-1)^{\frac{2m_{i}-\alpha_{i,4}\beta_{i,4}-s}{2}}\Gamma(\frac{s+\alpha_{i,4}\beta_{i,4}-2m_{i}}{2})$. We substitute the inner integral in  \eqref{eq:dgg_shadowng_pdf_2}, and apply the definition of Fox-H with the identity \cite[{eq. }1.59]{M-Foxh} to get \eqref{eq:dgg_shadowng_pdf}.
	\end{IEEEproof}

Finally we consider the effect of mobility in the last hop of RF-RIS i.e., from $(K_{2}-1)$ RIS to destination. 
\begin{my_proposition}
	The resultant  PDF of IG, dGG fading combined with mobility model for the last hop $g_{K_{2}}=g_{K_{2}}^{(l)} g_{K_{2}}^{(s)}g_{K_{2}}^{(f)}$ is given as
	\begin{align}\label{eq:comb_dgg_ph_mob}
	&f_{g_{K_{2}}^{}}(x) = \frac{6x^{-1}(m_{K_{2}}-1)^{m_{K_{2}}}}{\Gamma(\beta_{K_{2},3})\Gamma(\beta_{K_{2},4})} \nonumber \\ &H_{2,3}^{2,2} \bigg[\begin{array}{c} \frac{\zeta_{K_{2}}}{\sqrt{(m_{K_{2}}-1)^{}}} d_{K_{2}}^{\frac{a}{2}} x \end{array} \big\vert \begin{array}{c} (1+m_{K_{2}},\frac{1}{2}),(-1,\frac{a}{2}) \\ V_{1}\end{array}\bigg]
\end{align}	
where $\zeta_{K_{2}} = \big(\frac{\beta_{K_{2},4}}{\Omega_{K_{2},4}}\big)^{\frac{1}{\alpha_{K_{2},4}}} \big(\frac{\beta_{K_{2},3}}{\Omega_{K_{2},3}}\big)^{\frac{1}{\alpha_{K_{2},3}}}$ and $V_{1}=(\beta_{K_{2},4},\frac{1}{\alpha_{K_{2},4}}),(\beta_{K_{2},3},\frac{1}{\alpha_{K_{2},3}}),(-3,\frac{a}{2})$.
\end{my_proposition}
\begin{IEEEproof}
	The path loss component of last RF hop is $g_{K_{2}}^{(l)}=r^{-\frac{a}{2}}$, where the PDF of $r$ is given by \eqref{eq:mobility_power_pdf}. Now, the PDF of short-term fading combined with mobility model, $g_{K_{2}}=g_{K_{2}}^{(l)} g_{K_{2}}^{(s)}g_{K_{2}}^{(f)}$ can be computed as 
	\begin{eqnarray}\label{eq:comb_pdf}
		f_{g_{K_{2}}}(x) = \int_{0}^{d_{K_{2}}} f_{g_{K_{2}}^{}}(x/r) f_{r}(r) dr
	\end{eqnarray}
	And the PDF of $f_{g_{K_{2}}^{}}(x/r)$ for a given $r$ can be expressed as $f_{g_{K_{2}}^{}}(x/r) = r^{\frac{a}{2}} f_{g_{K_{2}}^{(sf)}}(x r^{\frac{a}{2}})$, we substitute it in \eqref{eq:comb_pdf} and apply the integral representation of Fox's H-function and interchange the order of integration to get
	\begin{eqnarray}\label{eq:comb_pdf_2}
		&\hspace{-4mm}f_{g_{K_{2}}}(x) = \frac{x^{-1}(m_{K_{2}}-1)^{m_{K_{2}}}}{\Gamma(\beta_{K_{2},3})\Gamma(\beta_{K_{2},4})} \frac{1}{2\pi \J} \int_{\mathcal{L}_{K_{2}}} (\frac{\zeta_{K_{2}}}{\sqrt{m_{K_{2}}-1}})^{n_{K_{2}}}  \nonumber \\ &\hspace{-2mm}\Gamma(\beta_{K_{2},4}-\frac{n_{K_{2}}}{\alpha_{K_{2},4}}) \Gamma(\beta_{K_{2},3}-\frac{n_{K_{2}}}{\alpha_{K_{2},3}}) \Gamma(-m_{K_{2}}+\frac{n_{K_{2}}}{2})\nonumber \\ & \hspace{-8mm}x^{n_{K_{2}}}(\int_{0}^{d_{K_{2}}} 6 r^{\frac{an_{K_{2}}}{2}}[\frac{r^{}}{d_{K_{2}}^{2}} - \frac{r^{2}}{d_{K_{2}}^{3}}] dr) dn_{K_{2}} \hspace{-4mm}
	\end{eqnarray}
	
	Now, solving the inner integral as $I = \frac{\Gamma(\frac{an_{K_{2}}}{2}+2)}{\Gamma(\frac{an_{K_{2}}}{2}+4)} d_{K_{2}}^{\frac{an_{K_{2}}}{2}}$. We substitute it back and use the integral  definition of Fox's H-function to get \eqref{eq:comb_dgg_ph_mob}.
\end{IEEEproof}

Next,  we develop a   framework to derive the PDF and CDF of cascaded channels $h$ and $g$, as presented in  Theorem 1:
\begin{my_theorem}\label{th:gen_prod_pdf_cdf}
	If  $X_{i}$, $i= 1 \cdots K$  denote  $K$ i.ni.d random variables with a PDF in the form of
	\begin{equation}\label{eq:gen_pdf}
		f_{X_{i}}(x) = \psi_{i} x^{\phi_{i}-1} H_{p,q}^{m,n} \bigg[\begin{array}{c}\zeta_{i} x \end{array} \big\vert \begin{array}{c}\{(a_{i,j},A_{i,j})\}_{j=1}^{p}\\ \{(b_{i,j},B_{i,j})\}_{j=1}^{q} \end{array}\bigg]
	\end{equation}
	then the PDF and CDF of the product of random variables $X=\prod_{i=1}^{K}X_{i}$ are given by
	\begin{align}
		&f_{X}(x) = \frac{1}{x} \prod_{i=1}^{K} \psi_{i} \zeta_{i}^{-\phi_{i}} H_{Kp,Kq}^{Km,Kn}\nonumber \\ &\bigg[\begin{array}{c}\prod_{i=1}^{K}\zeta_{i} x \end{array} \big \vert \begin{array}{c}\{\{(a_{i,j}+A_{i,j}\phi_{i},A_{i,j})\}_{j=1}^{p}\}_{i=1}^{K} \\ \{\{(b_{i,j}+B_{i,j}\phi_{i},B_{i,j})\}_{j=1}^{q}\}_{i=1}^{K} \end{array}\bigg]\label{eq:gen_prod_pdf}\\
		&F_{X}(x) = \prod_{i=1}^{K} \psi_{i} \zeta_{i}^{-\phi_{i}} H_{Kp+1,Kq+1}^{Km,Kn+1}\nonumber \\ &\hspace{-4mm}\bigg[\begin{array}{c}\prod_{i=1}^{K}\zeta_{i} x \end{array} \big \vert \begin{array}{c}(1,1),\{\{(a_{i,j}+A_{i,j}\phi_{i},A_{i,j})\}_{j=1}^{p}\}_{i=1}^{K} \\ \{\{(b_{i,j}+B_{i,j}\phi_{i},B_{i,j})\}_{j=1}^{q}\}_{i=1}^{K},(0,1) \end{array}\bigg]\label{eq:gen_prod_cdf}
	\end{align}
\end{my_theorem}
\normalsize
\begin{IEEEproof} 
	The proof is presented in Appendix A.
\end{IEEEproof}

\normalsize

Next, we use Theorem 1 to present the PDF and CDF of the cascaded  FSO and RF channels realized through multiple RIS modules:
\begin{my_corollary}
	If $h_{i}^{(tp)}$ is distributed according to \eqref{eq:dgg_pointing_error_pdf}, the PDF and CDF of cascaded FSO channel $h= \prod_{i=1}^{K_{1}}h_{i}^{(tp)}$ are
	\begin{align}\label{eq:pdf_prod_dgg_pointing}
		f_h(x) = &\frac{1}{x} \psi_{1} H_{K_{1},3K_{1}}^{3K_{1},0} \Big[\begin{array}{c} U_{1} x \end{array} \big\vert \begin{array}{c} \{(\rho_{i}^{2}+1,1)\}_{1}^{K_{1}} \\ V_{1}\end{array}\Big]
		\\ \label{eq:cdf_prod_dgg_pointing_error}
		F_{h}(x) =  &\psi_{1} H_{K_{1}+1,3K_{1}+1}^{3K_{1},1} \Big[\begin{array}{c} U_{1} x \end{array} \big\vert \begin{array}{c} (1,1),\{(\rho_{i}^{2}+1,1)\}_{1}^{K_{1}} \\ V_{1},(0,1)\end{array}\Big]
	\end{align}
\normalsize	
	where $\psi_{1} = \prod_{i=1}^{K_{1}} \frac{\rho_{i}^2}{\Gamma(\beta_{i,1})\Gamma(\beta_{i,2})}$,  $U_{1} = \prod_{i=1}^{K_{1}} \frac{1}{A_{0,i}} \big(\frac{\beta_{i,2}}{\Omega_{i,2}}\big)^{\frac{1}{\alpha_{i,2}}} \big(\frac{\beta_{i,1}}{\Omega_{i,1}}\big)^{\frac{1}{\alpha_{i,1}}}$ and $V_{1} = \{(\beta_{i,1},\frac{1}{\alpha_{i,1}}),(\beta_{i,2},\frac{1}{\alpha_{i,2}}),(\rho_{i}^{2},1)\}_{1}^{K_{1}}$.
\end{my_corollary}
\begin{IEEEproof}
	A direct application of  Theorem \ref{th:gen_prod_pdf_cdf} proves the Corollary 1.
\end{IEEEproof}
As a sanity check, to prove $\int_{0}^{\infty}f_{h}(x) \diff x=1$, we use the identity \cite[{eq. }2.8]{M-Foxh}
\begin{align}\label{eq:pdf_prod_dgg_pointing_proof}
	&\hspace{0mm}\int_{0}^{\infty} f_{h}(x) \diff x = \psi_{1} \prod_{i=1}^{K_{1}} \frac{\Gamma(\beta_{i,1}) \Gamma(\beta_{i,1})\Gamma(\rho_{i}^{2})}{\Gamma(\rho_{i}^{2}+1)}	= 1
\end{align}
\normalsize
\begin{my_corollary}
	The PDF and CDF of multi-hop RIS RF System which is the product of dGG combined with shadowing are given as
	\begin{align}\label{eq:pdf_prod_dgg}
		f_g(x) = &\frac{1}{x} \psi_{2} H_{K_{2}+1,2K_{2}+1}^{2K_{2},K_{2}+1} \Big[\begin{array}{c} U_{2} x \end{array} \big\vert \begin{array}{c} W_{2} \\ V_{2}\end{array}\Big]
		\\ \label{eq:cdf_prod_dgg}
		F_{g}(x) =  &\psi_{2} H_{K_{2}+2,2K_{2}+2}^{2K_{2},K_{2}+2} \Big[\begin{array}{c}  U_{2} x \end{array} \big\vert \begin{array}{c} W_{2},(1,1) \\ V_{2},(0,1)\end{array}\Big]
	\end{align}
	where $\psi_{2} = \prod_{i=1}^{K_{2}} \frac{(m_{i}-1)^{m_{i}}}{\Gamma(\beta_{i,3})\Gamma(\beta_{i,4})}$, $U_{2} = \prod_{i=1}^{K_{2}} \big(\frac{\beta_{i,4}}{\Omega_{i,4}}\big)^{\frac{1}{\alpha_{i,4}}}\frac{1}{\sqrt{(m_{i}-1)}}d_{K_{2}}^{\frac{a}{2}} \big(\frac{\beta_{i,3}}{\Omega_{i,3}}\big)^{\frac{1}{\alpha_{i,3}}}$, $V_{2} = \{(\beta_{i,3},\frac{1}{\alpha_{i,3}}),(\beta_{i,4},\frac{1}{\alpha_{i,4}})\}_{1}^{K_{2}},(-3,\frac{a}{2})$ and $W_{2}=\{(1+m_{i},\frac{1}{2})\}_{1}^{K_{2}}, \\(-1,\frac{a}{2})$.
\end{my_corollary}
\begin{IEEEproof}
	A direct application of Theorem \ref{th:gen_prod_pdf_cdf} completes the proof.
\end{IEEEproof}

 Using the IM/DD detector, the  SNR   of  the cascaded FSO  link is denoted by	$\gamma^{FSO}=\bar{\gamma}^{FSO} \lvert h \rvert^2$, and  the cascaded RF or R2V link  as  $\gamma^{R2V}=\gamma^{RF}+\gamma^{LOS}=\bar{\gamma}^{RF}\lvert g\rvert^{2}+\bar{\gamma}^{LOS}\lvert g_{LOS}\rvert^2$, where $\bar{\gamma}^{FSO}= \frac{P_1^{}\lvert h_{l}\rvert^{2}}{\sigma_{R}^2}$, $\bar{\gamma}^{RF}= \frac{P_2 |g_{l}|^2}{\sigma_{V}^2}$  and $\bar{\gamma}^{LOS}=\frac{P_2 |g_{LOS}^{(l)}|^2}{\sigma_{V}^2}$ are the SNR terms  without fading for the  FSO, RF and RF LOS links, respectively. Here, $h_{l}=\prod_{i=1}^{K_{1}}h_{i}^{(l)}$ and $g_{l}=\prod_{i=1}^{K_{2}}g_{i}^{(l)}$ are the overall path gain of the cascaded FSO links and RF links. 

The PDF of SNR of cascaded RF links and LOS link can be expressed using mathematical transformation as $f_{\gamma^{\rm RF}}(\gamma)=\frac{1}{2\sqrt{\bar{\gamma}^{\rm RF}\gamma}}f_{g_{\rm }^{}}(\sqrt{\frac{\gamma}{\bar{\gamma}^{\rm RF}}})$ and $  f_{\gamma^{\rm LOS}}(\gamma)=\frac{1}{2\sqrt{\bar{\gamma}^{\rm LOS}\gamma}}f_{g_{\rm LOS}^{(sf)}}(\sqrt{\frac{\gamma}{\bar{\gamma}^{\rm LOS}}})$ respectively.

We then compute the resultant PDF of RF SNR using inverse Laplace transform of moment generating function (MGF) of individual SNRs i.e., $f_{\gamma^{\rm R2V}}(\gamma)=\frac{1}{2\pi \J}\int_{L} M_{\gamma^{\rm RF}}(s) M_{\gamma^{\rm LOS}}(s) e^{s\gamma} ds$. Now, MGF of individual SNRs are computed as
\begin{eqnarray}\label{eq:mgf_snr_rf_ris_1}
&M_{\gamma^{\rm RF}}(s) = \int_{0}^{\infty} 	f_{\gamma^{\rm RF}}(\gamma) e^{-s\gamma} d\gamma \nonumber \\
&= \frac{\psi_{2}}{2}  \int_{0}^{\infty} \gamma^{-1} H_{K_{2}+1,2K_{2}+1}^{2K_{2},K_{2}+1} \nonumber \\&\Big[\begin{array}{c} U_{2} \sqrt{\frac{\gamma}{\bar{\gamma}^{\rm RF}}} \end{array} \big\vert \begin{array}{c} \{(1+m_{i},\frac{1}{2})\}_{1}^{K_{2}},(-1,\frac{a}{2}) \\ V_{2}\end{array}\Big] e^{-s\gamma} d\gamma
\end{eqnarray}	
Expanding the definition of Fox's H-function and solving the inner integral as $\int_{0}^{\infty} \gamma^{\frac{n_{1}}{2}-1} e^{-s\gamma} d\gamma = n_{}^{-\frac{n_{1}}{2}} \Gamma(\frac{n_{1}}{2})$. Substituting it back, we get
\begin{eqnarray}\label{eq:mgf_snr_rf_ris}
	&\hspace{-4mm}M_{\gamma^{\rm RF}}(s) = \frac{\psi_{2}}{2}   \frac{1}{2\pi \J} \int_{\mathcal{L}_{1}} (U_{2}\sqrt{\frac{1}{s_{}\bar{\gamma}^{\rm RF}}})^{n_{1}}  \Gamma(\frac{n_{1}}{2})\prod_{i=1}^{K_{2}}\nonumber \\ &\hspace{-4mm}\Gamma(\beta_{i,4}-\frac{n_{1}}{\alpha_{i,4}}) \Gamma(\beta_{i,3}-\frac{n_{1}}{\alpha_{i,3}}) \Gamma(-m_{i}+\frac{n_{1}}{2})  \frac{\Gamma(\frac{an_{1}}{2}+2)}{\Gamma(\frac{an_{1}}{2}+4)} dn_{1}
\end{eqnarray}
Similarly, MGF of SNR of RF LOS link is given as
\begin{eqnarray}\label{eq:mgf_snr_rf_los}
	&\hspace{-4mm}M_{\gamma^{\rm LOS}}(s) = \frac{\psi_{LOS}}{2}   \frac{1}{2\pi \J} \int_{\mathcal{L}_{2}} (\frac{U_{LOS}}{\sqrt{s_{}\bar{\gamma}^{\rm LOS}}})^{n_{2}}  \Gamma(-m_{LOS}+\frac{n_{LOS}}{2})\nonumber \\ &\hspace{-4mm}\Gamma(\frac{n_{2}}{2})\Gamma(\beta_{LOS,4}-\frac{n_{2}}{\alpha_{LOS,4}}) \Gamma(\beta_{LOS,3}-\frac{n_{2}}{\alpha_{LOS,3}})  dn_{2}
\end{eqnarray}
where $\psi_{LOS} =  \frac{1}{\Gamma(\beta_{LOS,3})\Gamma(\beta_{LOS,4})}$ and $U_{LOS} = \big(\frac{\beta_{LOS,4}}{\Omega_{LOS,4}}\big)^{\frac{1}{\alpha_{LOS,4}}}\frac{1}{\sqrt{(m_{LOS}-1)}} \big(\frac{\beta_{LOS,3}}{\Omega_{LOS,3}}\big)^{\frac{1}{\alpha_{LOS,3}}}$.

Now, the inner integral of PDF of resultant SNR becomes $\frac{1}{2\pi \J}\int_{L} s^{-\frac{n_{1}}{2}-\frac{n_{2}}{2}}e^{s\gamma} ds=\frac{ \gamma^{\frac{n_{1}}{2}+\frac{n_{2}}{2}-1} }{\Gamma(\frac{n_{1}}{2}+\frac{n_{2}}{2})}$. We substitute it back and apply the definition of bi-variate Fox's H-function, we get
\begin{eqnarray}\label{eq:pdf_snr_rf}
	&f_{\gamma^{\rm R2V}}(\gamma) = \frac{\psi_{2}\psi_{LOS}}{4} \gamma^{-1} H_{0,1:K_{2}+2,2K_{2}+1;2,2}^{0,0:2K_{2},K_{2}+2;2,2} \nonumber \\&\Big[\begin{array}{c} U_{2} \sqrt{\frac{\gamma}{\bar{\gamma}^{\rm RF}}}\\U_{LOS} \sqrt{\frac{\gamma}{\bar{\gamma}^{\rm LOS}}} \end{array} \big\vert \begin{array}{c} -:W_{2};W_{LOS} \\ (1:\frac{1}{2},\frac{1}{2}):V_{2}:V_{LOS}\end{array}\Big]
\end{eqnarray}
where $W_{2}=\{(1+m_{i},\frac{1}{2})\}_{1}^{K_{2}},(-1,\frac{a}{2}),(1,\frac{1}{2})$, $V_{2} = \{(\beta_{i,3},\frac{1}{\alpha_{i,3}}),(\beta_{i,4},\frac{1}{\alpha_{i,4}})\}_{1}^{K_{2}},(-3,\frac{a}{2})$, $W_{LOS}=(1+m_{LOS},\frac{1}{2}),(1,\frac{1}{2})$ and $V_{LOS}=(\beta_{LOS,3},\frac{1}{\alpha_{LOS,3}}),(\beta_{LOS,4},\frac{1}{\alpha_{LOS,4}})$.

Similarly, CDF of the resultant SNR can be obtained as $F_{\gamma^{\rm R2V}}(\gamma)=\frac{1}{2\pi \J}\int_{L} M_{\gamma^{\rm RF}}(s) M_{\gamma^{\rm LOS}}(s) s^{-1} e^{s\gamma} ds$
\begin{eqnarray}\label{eq:cdf_snr_rf}
	&F_{\gamma^{\rm R2V}}(\gamma) = \frac{\psi_{2}\psi_{LOS}}{4}  H_{0,1:K_{2}+2,2K_{2}+1;2,2}^{0,0:2K_{2},K_{2}+2;2,2} \nonumber \\&\Big[\begin{array}{c} U_{2} \sqrt{\frac{\gamma}{\bar{\gamma}^{\rm RF}}}\\U_{LOS} \sqrt{\frac{\gamma}{\bar{\gamma}^{\rm LOS}}} \end{array} \big\vert \begin{array}{c} -:W_{2};W_{LOS} \\ (0:\frac{1}{2},\frac{1}{2}):V_{2}:V_{LOS}\end{array}\Big]
\end{eqnarray}

In the following section, we analyze the system performance employing fixed-gain AF relaying to integrate the FSO link with  vehicular transmissions over RF.

\section{Fixed-Gain AF Relaying for Mixed System}
In this section, we analyze the mixed system by  employing the fixed-gain AF relaying protocol to the received FSO signal   forwarding the signal over RF to the desired vehicle.  In contrast to the channel-assisted AF relaying,  gain applied at the relay in the fixed-gain relaying protocol  is computed using  received signal from the source to the relay.  The fixed-gain AF relaying is relatively simpler in terms of computational complexity due to the relaxation on the knowledge of  the perfect channel state information. The resultant SNR  of the mixed system comprising of  FSO and RF links can be expressed as \cite{Hasna_2004_AF}:
\begin{equation}
	\label{eq:af}
	{\gamma^{AF}} = \frac{\gamma^{FSO}\gamma^{R2V}}{\gamma^{R2V}+C}
\end{equation}
where $C={P_2}/G^2\sigma_{w_2}^2$. Applying the theory of random variables on \eqref{eq:af}, the PDF of ${\gamma^{AF}}$ is: 
\begin{equation} \label{eq:snr_pdf_eqn_af_gen}
	f_{\gamma^{AF}}(\gamma) = \int_{0}^{\infty} {f_{\gamma^{FSO}}\left(\frac{\gamma(x + C)}{x}\right)} {f_{\gamma^{R2V}}(x)} \frac{x + C}{{x}} {dx}
\end{equation}  
where $f_{\gamma^{FSO}}(\gamma)$ and $f_{\gamma^{R2V}}(\gamma)$ are the PDF of SNR for FSO and RF links, respectively. We use  \eqref{eq:pdf_prod_dgg_pointing} and \eqref{eq:pdf_prod_dgg} with a simple transformation of random variable to get the PDF of SNR for FSO $f_{\gamma^{FSO}}(\gamma) = \frac{1}{2\sqrt{\gamma\bar{\gamma}^{FSO}}}f_{h}\big(\sqrt{\frac{\gamma}{\bar{\gamma}^{FSO}}}\big)$ and RF link $f_{\gamma^{R2V}}(\gamma) = \frac{1}{2\sqrt{\gamma\bar{\gamma}^{R2V}}}f_{g}\big(\sqrt{\frac{\gamma}{\bar{\gamma}^{R2V}}}\big)$.

\begin{my_lemma}\label{lemma:pdfcdf_af}
	The PDF and CDF of the fixed-gain AF relaying system in \eqref{eq:snr_pdf_eqn_af_gen}  is given as
\begin{align} 
	&f_{\gamma^{AF}}(\gamma) = \frac{1}{8\gamma} \psi_{1} \psi_{2} \psi_{LOS} H_{1,0:3K_{1},K_{1}+1;K_{2}+2,2K_{2}+1;2,2}^{0,1:0,3K_{1};2K_{2},K_{2}+2;2,2} \nonumber \\ &\Bigg[\begin{array}{c} U_{1}^{-1} \sqrt{\frac{\bar{\gamma}^{FSO}}{\gamma}} \\ U_{2} \sqrt{\frac{C}{\bar{\gamma}^{RF}}} \\ U_{LOS} \sqrt{\frac{C}{\bar{\gamma}^{LOS}}} \end{array} \big\vert \begin{array}{c} (1:\frac{1}{2},\frac{1}{2},\frac{1}{2}):V_{1};W_{2};W_{LOS}\\ - : W_{1},(1,\frac{1}{2}) ; V_{2};V_{LOS}\end{array} \Bigg]\label{eq:snr_pdf_eqn_af}\\
	&\hspace{0mm}F_{\gamma^{AF}}(\gamma) = \frac{1}{8} \psi_{1} \psi_{2} \psi_{LOS} H_{1,0:3K_{1}+1,K_{1}+2;K_{2}+2,2K_{2}+1;2,2}^{0,1;1,3K_{1};2K_{2},K_{2}+2;2,2} \nonumber \\ &\hspace{0mm}\Bigg[\begin{array}{c} U_{1}^{-1} \sqrt{\frac{\bar{\gamma}^{FSO}}{\gamma}} \\ U_{2} \sqrt{\frac{C}{\bar{\gamma}^{RF}}} \\ U_{LOS} \sqrt{\frac{C}{\bar{\gamma}^{LOS}}} \end{array} \big\vert \begin{array}{c} (1:\frac{1}{2},\frac{1}{2},\frac{1}{2}):V_{1},(1,\frac{1}{2}); W_{2};W_{LOS}\\ - : (0,\frac{1}{2}),W_{1},(1,\frac{1}{2}) ; V_{2};V_{LOS}\end{array} \Bigg]\label{eq:snr_cdf_eqn_af}
\end{align}
\normalsize 
where $V_{1}=\{(1-\beta_{i,1},\frac{1}{\alpha_{i,1}}),(1-\beta_{i,2},\frac{1}{\alpha_{i,2}}),(1-\rho_{i}^{2},1)\}_{1}^{K_{1}}$, $W_{1}=\{(-\rho_{i}^{2},1)\}_{1}^{K_{1}}$,  $V_{2}=\{(\beta_{i,3},\frac{1}{\alpha_{i,3}}),(\beta_{i,4},\frac{1}{\alpha_{i,4}})\}_{1}^{K_{2}},(-3,\frac{a}{2})$, $W_{2}=\{(1+m_{i},\frac{1}{2})\}_{1}^{K_{2}},(-1,\frac{a}{2}),(1,\frac{1}{2})$, $W_{LOS}=(1+m_{LOS},\frac{1}{2}),(1,\frac{1}{2})$ and $V_{LOS}=(\beta_{LOS,3},\frac{1}{\alpha_{LOS,3}}),(\beta_{LOS,4},\frac{1}{\alpha_{LOS,4}})$.	
\end{my_lemma}

\begin{IEEEproof} We substitute PDF of FSO and RF links in \eqref{eq:snr_pdf_eqn_af_gen}, use  the integral representation of the Fox's-H function and interchanging the order of integration to express
\begin{align} \label{eq:snr_pdf_eqn_af_1}
	&f_{\gamma^{AF}}(\gamma) = \frac{\psi_{1}\psi_{2} \psi_{LOS}}{8\gamma}   \frac{1}{\big(2\pi \J\big)^{3}} \int_{\mathcal{L}_{1}} \int_{\mathcal{L}_{2}} \int_{\mathcal{L}_{3}} \big(U_{1} \sqrt{\frac{\gamma}{\bar{\gamma}^{FSO}}}\big)^{-n_{1}}  \nonumber \\ & \prod_{i=1}^{K_{1}} \Gamma(\beta_{i,2}+\frac{n_{1}}{\alpha_{i,2}})\Gamma(\beta_{i,1}+\frac{n_{1}}{\alpha_{i,1}}) \frac{\Gamma(\rho_{i}^{2}+n_{1})}{\Gamma(\rho_{i}^{2}+n_{1}+1)} \nonumber \\ &\bigg(U_{2} \sqrt{\frac{1}{\bar{\gamma}^{RF}}}\bigg)^{-n_{2}}  \prod_{i=1}^{K_{2}} \Gamma(\beta_{i,4}+\frac{n_{2}}{\alpha_{i,4}}) \Gamma(\beta_{i,3}+\frac{n_{2}}{\alpha_{i,3}})  \nonumber \\ &\Gamma(-m_{i}-\frac{n_{2}}{2})  \Gamma(-\frac{n_{2}}{2})\frac{\Gamma(-\frac{an_{2}}{2}+2)}{\Gamma(\frac{-an_{2}}{2}+4)}\bigg(U_{LOS} \sqrt{\frac{1}{\bar{\gamma}^{LOS}}}\bigg)^{-n_{3}}  \nonumber\\&\Gamma(\beta_{LOS,4}+\frac{n_{3}}{\alpha_{LOS,4}}) \Gamma(\beta_{LOS,3}+\frac{n_{3}}{\alpha_{LOS,3}})\Gamma(-m_{LOS}-\frac{n_{3}}{2})  \nonumber\\&\frac{\Gamma(-\frac{n_{3}}{2})}{\Gamma(\frac{n_{2}}{2}+\frac{n_{3}}{2})}\big(\int_{0}^{\infty} x^{-1-\frac{n_{2}}{2}-\frac{n_{3}}{2}} \big(\frac{x+C}{x}\big)^{-\frac{n_{1}}{2}} \diff x\big) \diff n_{1} \diff n_{2}  \diff n_{3}
\end{align}

We use the identities \cite[(3.194/3)]{integrals}  and \cite[(8.384/1)]{integrals} to express inner integral   in terms of Gamma function:
\begin{align} \label{eq:snr_pdf_eqn_af_2}
	&\int_{0}^{\infty} x^{-1-\frac{n_{2}}{2}-\frac{n_{3}}{2}} \big(\frac{x+C}{x}\big)^{-\frac{n_{1}}{2}} \diff x =\nonumber\\& \frac{C^{-\frac{n_{2}}{2}-\frac{n_{3}}{2}}\Gamma(\frac{n_{2}}{2}+\frac{n_{3}}{2})\Gamma(-\frac{n_{2}}{2}-\frac{n_{3}}{2}+\frac{n_{1}}{2})}{\Gamma(\frac{n_{1}}{2})}
\end{align}

Substituting \eqref{eq:snr_pdf_eqn_af_2} in \eqref{eq:snr_pdf_eqn_af_1} and  applying the definition of trivariate Fox's-H function, we get \eqref{eq:snr_pdf_eqn_af}.

To develop an analytical expression for the CDF, we use  \eqref{eq:snr_pdf_eqn_af} in $F_{\gamma^{AF}}(\gamma)=\int_0^\gamma f_{\gamma^{AF}}(t) \diff t$,  apply the definition of Fox's-H function to solve  $\int_{0}^{\gamma} t^{\frac{n_{1}}{2}-1}\diff t={\gamma^{\frac{n_{1}}{2}}}{\frac{2}{n_{1}}}=\gamma^{\frac{n_{1}}{2}}\frac{\Gamma(\frac{n_{1}}{2})}{\Gamma(1+\frac{n_{1}}{2})}$. and then use the definition of trivariate Fox's-H function, to get \eqref{eq:snr_cdf_eqn_af}, which completes the proof of Lemma.

\end{IEEEproof}
Note that trivariate Fox's H-function has been extensively used for the representation of complicated fading distribution functions. Moreover, numerical computational codes are available  in MATLAB  for the computation of  trivariate Fox's H-function.  

\normalsize
Next, we use the results of Lemma \ref {lemma:pdfcdf_af} to present analytical expressions of outage probability, and average BER of the integrated FSO-RF  system equipped with multiple RIS.

\begin{figure*}
	\begin{align}\label{eq:pout_asymp_af}
		&P_{\rm out}^{AF,\infty} \approx \frac{\psi_{1} \psi_{2} \psi_{LOS}}{8} \Gamma(\frac{p_{2}}{2}+\frac{p_{3}}{2}+\frac{p_{1}}{2}) \bigg[\bigg(U_{1}^{-1} \sqrt{\frac{\bar{\gamma}^{FSO}}{\gamma_{th}}}\bigg)^{-p_{1}} \prod_{i=1}^{K_{1}} \frac{2}{p_{1}}  \Gamma(\beta_{i,2}-\frac{p_{1}}{\alpha_{i,2}})\Gamma(\beta_{i,1}-\frac{p_{1}}{\alpha_{i,1}})  \frac{\Gamma(\rho_{i}^{2}-p_{1})}{\Gamma(\rho_{i}^{2}-p_{1}+1)\Gamma(-\frac{p_{1}}{2})}\bigg]     \nonumber \\ &\bigg[ \bigg(U_{2} \sqrt{\frac{C}{\bar{\gamma}^{RF}}}\bigg)^{p_{2}}   \prod_{i=1}^{K_{2}} \Gamma(\beta_{i,4}-\frac{p_{2}}{\alpha_{i,4}})\Gamma(\beta_{i,3}-\frac{p_{2}}{\alpha_{i,3}}) \Gamma(\frac{p_{2}}{2}) \Gamma(-m_{i}+\frac{p_{2}}{2})  \frac{\Gamma(\frac{ap_{2}}{2}+2)}{\Gamma(\frac{ap_{2}}{2}+4)} \nonumber \\  &\bigg(U_{LOS} \sqrt{\frac{C}{\bar{\gamma}^{LOS}}}\bigg)^{p_{3}}   \Gamma(\frac{p_{3}}{2}) \Gamma(\beta_{LOS,4}-\frac{p_{3}}{\alpha_{LOS,4}}) \Gamma(\beta_{LOS,3}-\frac{p_{3}}{\alpha_{LOS,3}}) \Gamma(-m_{LOS}+\frac{p_{3}}{2})  \bigg]
	\end{align}
	\hrule
\end{figure*}

\subsection{Outage Probability}\label{sec:outage_probability_af}
The outage probability of a system is a probabilistic measure of the instantaneous  SNR falling below certain threshold SNR $\gamma_{\rm th}$ i.e., $ P_{\rm out}=Pr(\gamma \le \gamma_{\rm th})=F_{\gamma}(\gamma_{th})$. Thus, an exact outage probability for the mixed FSO-RF system can be simply expressed using $\gamma=\gamma_{\rm th}$ in the CDF formulation $F_{\gamma^{AF}}(\gamma)$ of \eqref{eq:snr_cdf_eqn_af}.

 We use \cite{Kilbas_FoxH} to compute  the residue of the  trivariate Fox's H-function at the dominant poles of \eqref{eq:snr_cdf_eqn_af} to develop the asymptotic expression in the high SNR regime, as given in \eqref{eq:pout_asymp_af}.  The first term of \eqref{eq:pout_asymp_af} corresponds to the residue at dominant pole of FSO coefficients represented by $p_{1}=\min{\{\{\frac{\beta_{i,1}\alpha_{i,1}}{2},\frac{\beta_{i,2}\alpha_{i,2}}{2},\frac{\rho_{i}^{2}}{2}\}_{1}^{K_{1}}\}}$ and later terms represent residues corresponding to RF coefficients at dominant poles $p_{2}=\min{\{\{\frac{\beta_{i,3}\alpha_{i,3}}{2},\frac{\beta_{i,4}\alpha_{i,4}}{2}\}_{1}^{K_{2}}\}}$ and $p_{3}=\min{\{\frac{\beta_{LOS,3}\alpha_{LOS,3}}{2},\frac{\beta_{LOS,4}\alpha_{LOS,4}}{2}\}}$. To get the diversity order $G_{\rm out}$ of the considered system, we express \eqref{eq:pout_asymp_af} as $P_{\rm out}^{AF,\infty} \propto \gamma_{0}^{-G_{\rm out}}$. Finally,  selection of dominant poles in \eqref{eq:pout_asymp_af}  gives rise to the outage diversity order of the systems as $G_{\rm out} = p_{1}+p_{2}+p_{3}$. The derived diversity order depicts that the performance of the considered multi-hop transmission depends on the link with the minimum of channel parameters. Thus, a performance degradation due to the channel fading and  pointing errors can be compensated with an increase in the number of RISs for a given distance in both links.

\subsection{Average BER}\label{sec:ber_af}
Using the CDF of the SNR, the average BER of a communication system   is given by \cite{Ansari2011_ber}:
\begin{equation} \label{eq:ber}
	\bar{P}_{e} = \frac{q^p}{2\Gamma(p)}\int_{0}^{\infty} \gamma^{p-1} {e^{{-q \gamma}}} F_{\gamma} (\gamma)   d\gamma
\end{equation}
where $p$ and $q$ parameterize different modulation schemes. 

We substitute \eqref{eq:snr_cdf_eqn_af} in \eqref{eq:ber} and use the trivariate Fox's-H function \cite{M-Foxh} definition, express the inner integral  as $\int_{0}^{\infty} \gamma^{p-\frac{n_{1}}{2}-1} {e^{{-q \gamma}}}  d\gamma = \frac{\Gamma(p-\frac{n_{1}}{2})}{q^{p-\frac{n_{1}}{2}}}$, and again apply the definition of trivariate Fox's-H function to express the average BER of the integrated system as

\begin{align} 
	&\bar{P_e^{AF}} =  \frac{1}{16\Gamma(p)} \psi_{1} \psi_{2} \psi_{LOS} H_{1,0:3K_{1}+1,K_{1}+3;K_{2}+2,2K_{2}+1;2,2}^{0,1;2,3K_{1};2K_{2},K_{2}+2;2,2} \nonumber \\ &\hspace{0mm}\Bigg[\begin{array}{c} U_{1}^{-1} \sqrt{q\bar{\gamma}^{FSO}} \\ U_{2} \sqrt{\frac{C}{\bar{\gamma}^{RF}}} \\ U_{LOS} \sqrt{\frac{C}{\bar{\gamma}^{LOS}}} \end{array} \big\vert \begin{array}{c} (1:\frac{1}{2},\frac{1}{2},\frac{1}{2}):V_{1}; W_{2};W_{LOS}\\ - : W_{1} ; V_{2};V_{LOS}\end{array} \Bigg]\label{eq:ber_af}
\end{align}
\normalsize 
where $V_{1}=\{(1-\beta_{i,1},\frac{1}{\alpha_{i,1}}),(1-\beta_{i,2},\frac{1}{\alpha_{i,2}}),(1-\rho_{i}^{2},1)\}_{1}^{K_{1}},(1,\frac{1}{2})$, $W_{1}=(p,\frac{1}{2}),(0,\frac{1}{2}),\{(-\rho_{i}^{2},1)\}_{1}^{K_{1}},(1,\frac{1}{2})$,  $V_{2}=\{(\beta_{i,3},\frac{1}{\alpha_{i,3}}),(\beta_{i,4},\frac{1}{\alpha_{i,4}})\}_{1}^{K_{2}},(-3,\frac{a}{2})$, $W_{2}=\{(1+m_{i},\frac{1}{2})\}_{1}^{K_{2}},(-1,\frac{a}{2}),(1,\frac{1}{2})$, $W_{LOS}=(1+m_{LOS},\frac{1}{2}),(1,\frac{1}{2})$ and $V_{LOS}=(\beta_{LOS,3},\frac{1}{\alpha_{LOS,3}}),(\beta_{LOS,4},\frac{1}{\alpha_{LOS,4}})$.

Note that an asymptotic expression for the average BER can be derived using the similar procedure applied for the outage probability (expressions are not listed due to the space constraint).

Similarly, ergodic capacity for the mixed system  can be derived applying  mathematical procedure as developed through average BER derivation (omitted here because of the page constraints).

\section{DF Relaying for Mixed System}
Although the fixed-gain AF relaying is a simple technique to implement, it is desirable to analyze the performance of the near-optimal DF protocol as a benchmark for the system performance. 

Since $\gamma^{FSO}$ and $\gamma^{R2V}$ are independent, the end-to-end SNR of DF relaying system is given as $\gamma=\min\{\gamma^{FSO},\gamma^{R2V}\}$. Hence, the CDF of the SNR is
\begin{equation}\label{eq:cdf_snr_df}
	F^{DF}_{\gamma}(\gamma) = F_{\gamma^{FSO}}(\gamma)+F_{\gamma^{R2V}}(\gamma)-F_{\gamma^{FSO}}(\gamma)F_{\gamma^{R2V}}(\gamma)
\end{equation}

\begin{figure*}
	\begin{align}\label{eq:pout_aymp}
		&P_{\rm{out}}^{\infty} \approx \frac{\psi_{1}}{2} \bigg[\bigg(U_{1}^{-1} \sqrt{\frac{\bar{\gamma}^{FSO}}{\gamma_{th}}}\bigg)^{-p_{1}} \prod_{i=1}^{K_{1}} \frac{2}{p_{1}}  \Gamma(\beta_{i,2}-\frac{p_{1}}{\alpha_{i,2}})\Gamma(\beta_{i,1}-\frac{p_{1}}{\alpha_{i,1}})  \frac{\Gamma(\rho_{i}^{2}-p_{1})}{\Gamma(\rho_{i}^{2}-p_{1}+1)\Gamma(-\frac{p_{1}}{2})}\bigg]   +  \nonumber \\ &\frac{\psi_{2} \psi_{LOS}}{4} \Gamma(\frac{p_{2}}{2}+\frac{p_{3}}{2})\bigg[ \bigg(U_{2} \sqrt{\frac{C}{\bar{\gamma}^{RF}}}\bigg)^{p_{2}}   \prod_{i=1}^{K_{2}} \Gamma(\beta_{i,4}-\frac{p_{2}}{\alpha_{i,4}})\Gamma(\beta_{i,3}-\frac{p_{2}}{\alpha_{i,3}}) \Gamma(\frac{p_{2}}{2}) \Gamma(-m_{i}+\frac{p_{2}}{2})  \frac{\Gamma(\frac{ap_{2}}{2}+2)}{\Gamma(\frac{ap_{2}}{2}+4)} \nonumber \\  &\bigg(U_{LOS} \sqrt{\frac{C}{\bar{\gamma}^{LOS}}}\bigg)^{p_{3}}   \Gamma(\frac{p_{3}}{2}) \Gamma(\beta_{LOS,4}-\frac{p_{3}}{\alpha_{LOS,4}}) \Gamma(\beta_{LOS,3}-\frac{p_{3}}{\alpha_{LOS,3}}) \Gamma(-m_{LOS}+\frac{p_{3}}{2})  \bigg]
	\end{align}
	\hrule
\end{figure*}
\subsection{Outage Probability}\label{sec:outage_probability}
Similar to the fixed-gain, an exact outage probability for the integrated FSO-RF system can be expressed using the CDF at a specific threshold value of SNR	$P^{DF}_{\rm out} = F^{DF}_{\gamma}(\gamma_{\rm th})$. Further, we use the series expansion of  Fox's H-function  \cite[Th. 1.11]{Kilbas_FoxH} to represent  the outage probability in the high SNR regime, as given in \eqref{eq:pout_aymp}. 

\subsection{Average BER}\label{sec:ber}
For the DF based dual-hop FSO-RF system, the average BER can be expressed using the average BER of individual links \cite{Tsiftsis2006}:
\begin{equation}\label{eq:ber_df_1}
	\bar{P_{e}}^{} = \bar{P_{e}}^{(FSO)}+\bar{P_{e}}^{(R2V)}-2\bar{P_{e}}^{(FSO)}\bar{P_{e}}^{(R2V)}
\end{equation}
where $\bar{P_{e}}^{(FSO)}$ and $\bar{P_{e}}^{(R2V)}$ are average BER of the cascaded FSO and cascaded RF links,  respectively.

To derive $\bar{P_{e}}^{(FSO)}$, we substitute $F_{\gamma^{FSO}}(\gamma)$ in \eqref{eq:ber},  expand the definition of Fox's-H function and  interchange the order of integration to solve the inner integral $\int_{0}^{\infty} \gamma^{p+\frac{n_{1}}{2}-1} {e^{{-q \gamma}}}  d\gamma = \frac{\Gamma(p+\frac{n_{1}}{2})}{q^{p+\frac{n_{1}}{2}}}$. We then apply the definition of Fox's-H function to get
\begin{eqnarray}\label{eq:ber_mult_fso_2}
	\bar{P_{e}}^{(FSO)} = \frac{\psi_{1}}{2\Gamma(p)}  H_{K_{1}+2,3K_{1}+1}^{3K_{1},2} \Big[\begin{array}{c}    \frac{U_{1}}{\sqrt{q\bar{\gamma}^{FSO}}} \end{array} \big\vert \begin{array}{c} (1,1),V \\ V_{1},(0,1)\end{array}\Big]
\end{eqnarray}
where $\psi_{1} = \prod_{i=1}^{K_{1}} \frac{\rho_{i}^2}{\Gamma(\beta_{i,1})\Gamma(\beta_{i,2})}$, $U_{1} =\prod_{i=1}^{K_{1}}\frac{1}{A_{0,i}} \big(\frac{\beta_{i,2}}{\Omega_{i,2}}\big)^{\frac{1}{\alpha_{i,2}}} \big(\frac{\beta_{i,1}}{\Omega_{i,1}}\big)^{\frac{1}{\alpha_{i,1}}}$, $V=(1-p,\frac{1}{2}),\{(\rho_{i}^{2}+1,1)\}_{1}^{K_{1}}$ and $V_{1} = \{(\beta_{i,1},\frac{1}{\alpha_{i,1}}),(\beta_{i,2},\frac{1}{\alpha_{i,2}}),(\rho_{i}^{2},1)\}_{1}^{K_{1}}$. 

Similarly, the average BER of the RF link is:
\begin{eqnarray}\label{eq:ber_mult_rf_2}
	&\bar{P_{e}}^{(R2V)} = \frac{1}{8\Gamma(p)} \psi_{2}\psi_{LOS} H_{1,1:K_{2}+2,2K_{2}+1;2,2}^{0,1:2K_{2},K_{2}+2;2,2} \nonumber \\&\Big[\begin{array}{c} U_{2} \frac{1}{\sqrt{q\bar{\gamma}^{\rm RF}}}\\U_{LOS} \frac{1}{\sqrt{q\bar{\gamma}^{\rm LOS}}} \end{array} \big\vert \begin{array}{c} (1-p,\frac{1}{2},\frac{1}{2}):W_{2};W_{LOS} \\ (0:\frac{1}{2},\frac{1}{2}):V_{2}:V_{LOS}\end{array}\Big]
\end{eqnarray}
where $W_{2}=\{(1+m_{i},\frac{1}{2})\}_{1}^{K_{2}},(-1,\frac{a}{2}),(1,\frac{1}{2})$, $V_{2} = \{(\beta_{i,3},\frac{1}{\alpha_{i,3}}),(\beta_{i,4},\frac{1}{\alpha_{i,4}})\}_{1}^{K_{2}},(-3,\frac{a}{2})$, $W_{LOS}=(1+m_{LOS},\frac{1}{2}),(1,\frac{1}{2})$ and $V_{LOS}=(\beta_{LOS,3},\frac{1}{\alpha_{LOS,3}}),(\beta_{LOS,4},\frac{1}{\alpha_{LOS,4}})$.

The asymptotic expressions for average BER can be derived applying the  similar procedure as adopted for  the outage probability.

\begin{figure*}[!htbp]
	\centering
	\subfigure[Outage probability.]{\includegraphics[scale=0.35]{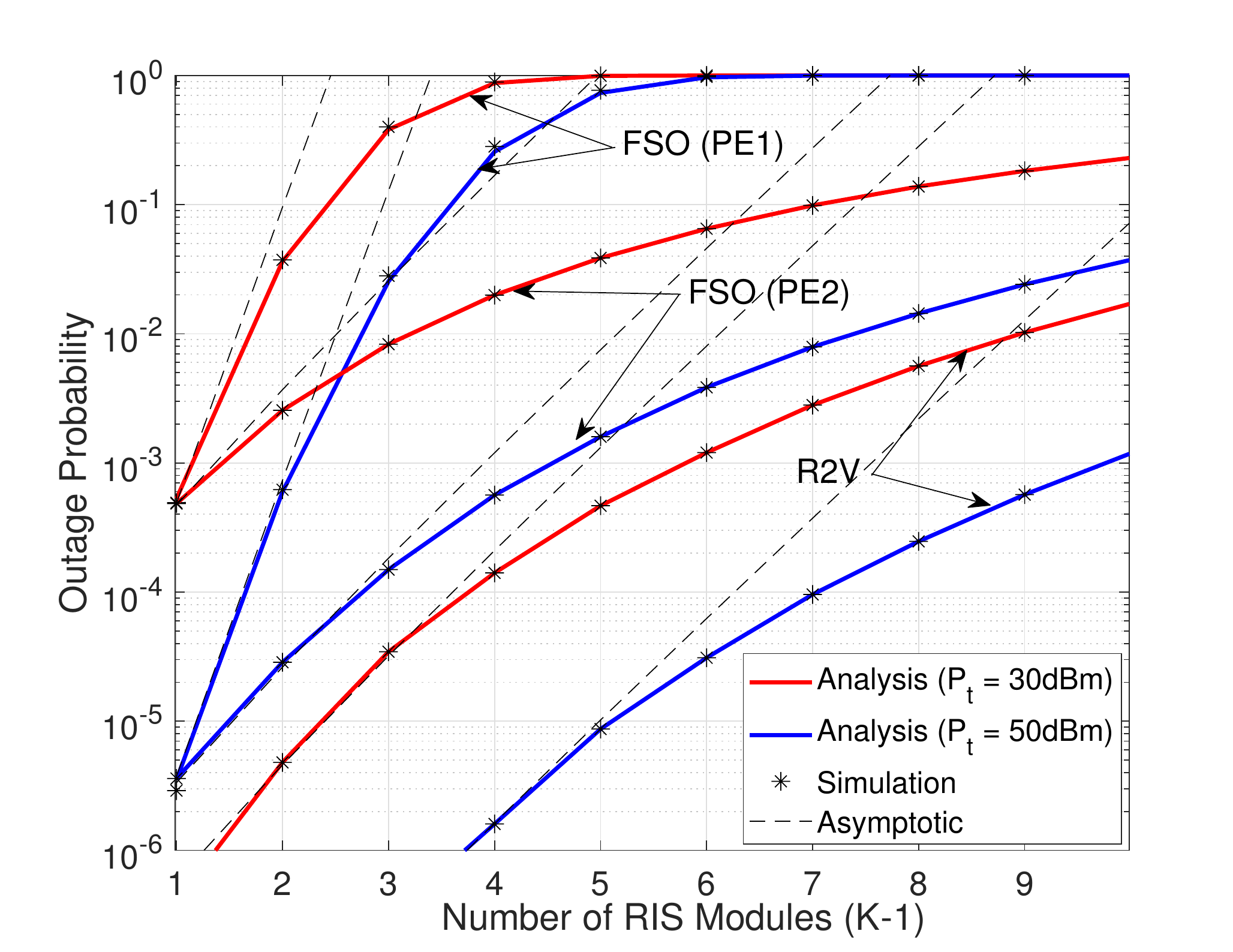}} \hspace{-6mm}
	\subfigure[Average BER. ]{\includegraphics[scale=0.35]{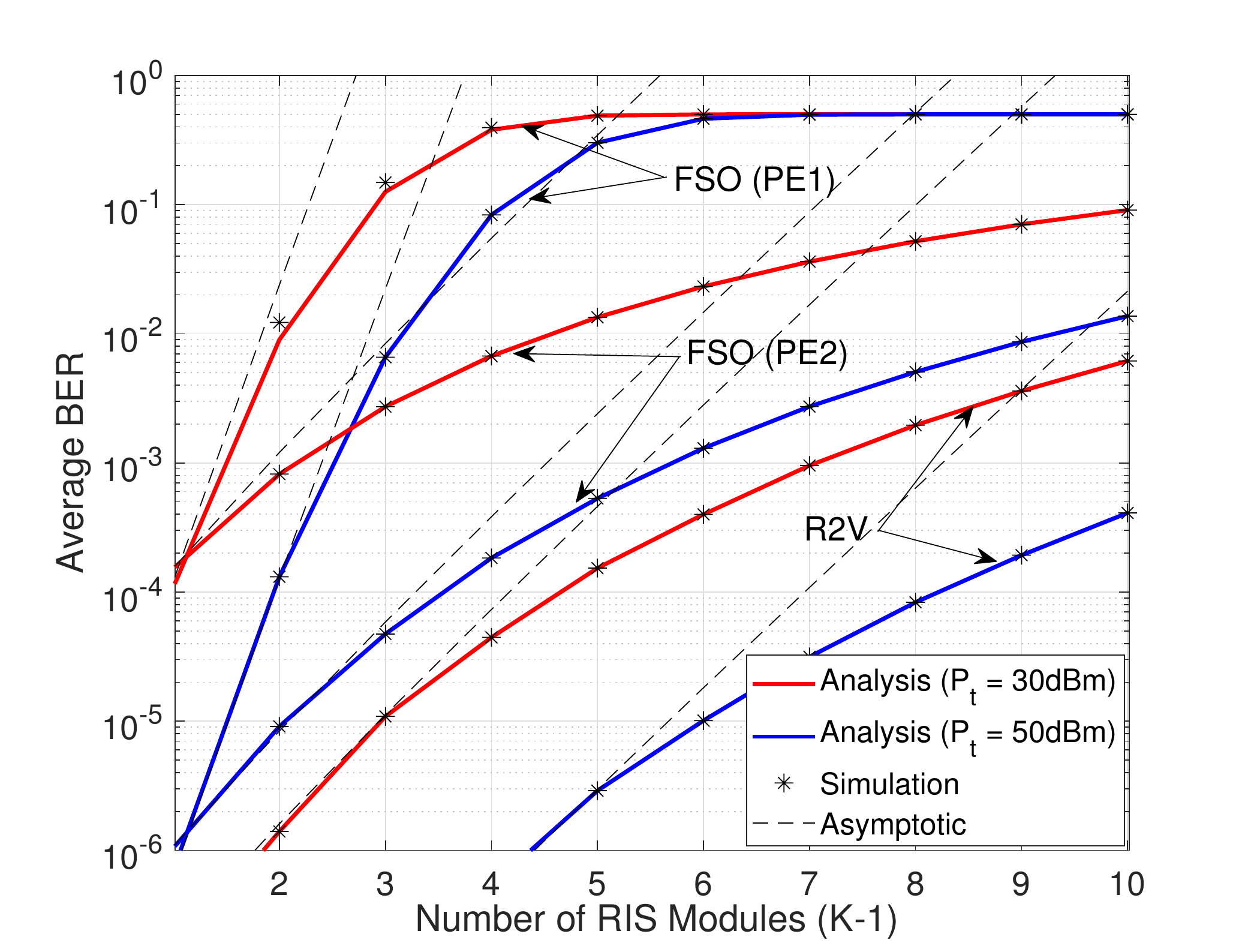}} \hspace{-6mm}
	\caption{Performance demonstration of multi-RIS transmissions for FSO and  R2V links ($K_{1}=K_{2}=K$) under similar channel conditions.}
	\label{fig:outage}
	\label{fig:ber}
\end{figure*}

\begin{figure*}[!htbp]
	\centering
	\subfigure[Outage probability.]{\includegraphics[scale=0.35]{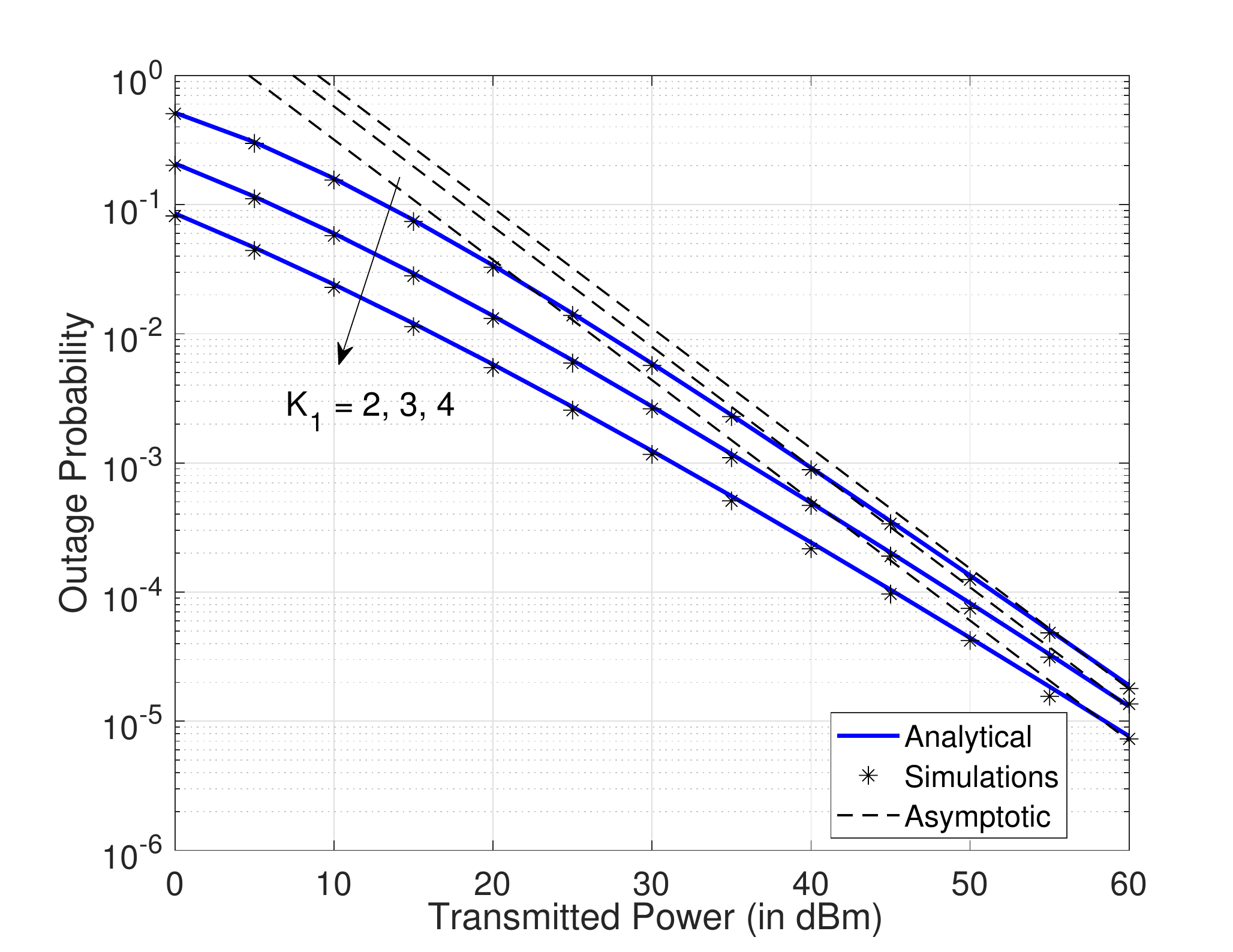}} \hspace{-6mm}
	\subfigure[Average BER. ]{\includegraphics[scale=0.35]{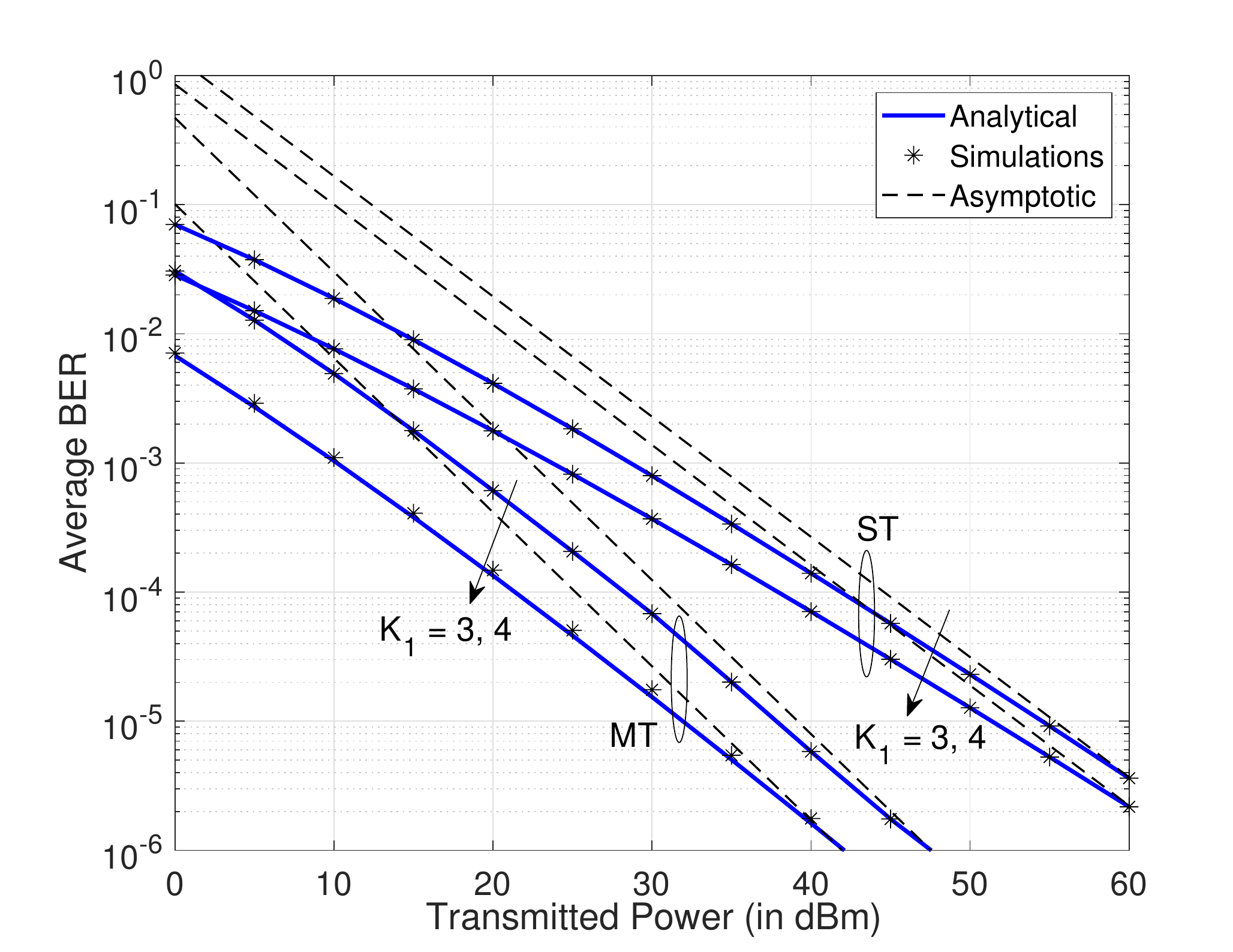}} \hspace{-6mm}
	\caption{Performance demonstration of multi-RIS transmissions for FSO link.}
	\label{fig:fso_outage}
	\label{fig:fso_ber}
\end{figure*}

\begin{figure*}[!htbp]
	\centering
	\subfigure[Outage probability.]{\includegraphics[scale=0.35]{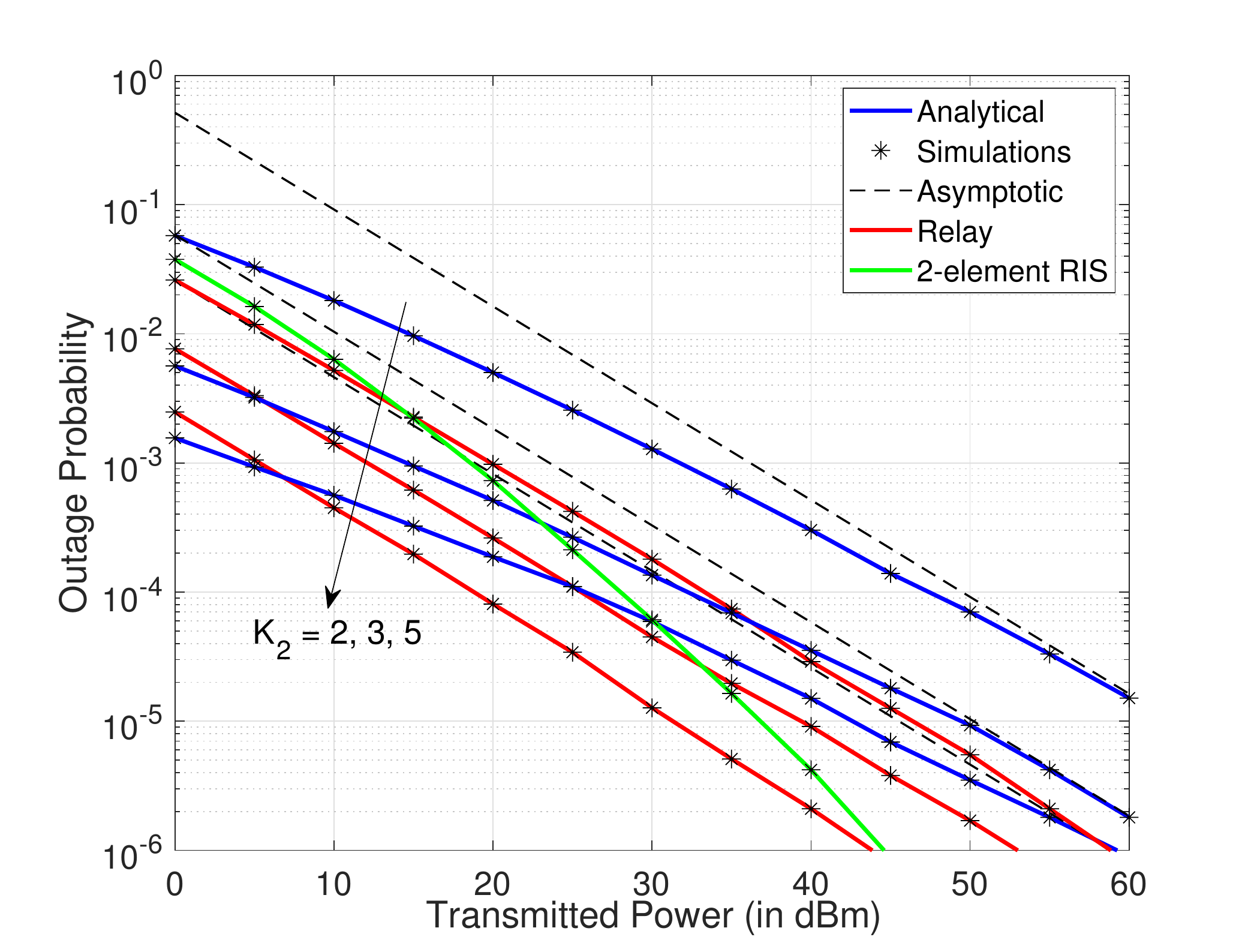}} \hspace{-6mm}
	\subfigure[Average BER. ]{\includegraphics[scale=0.35]{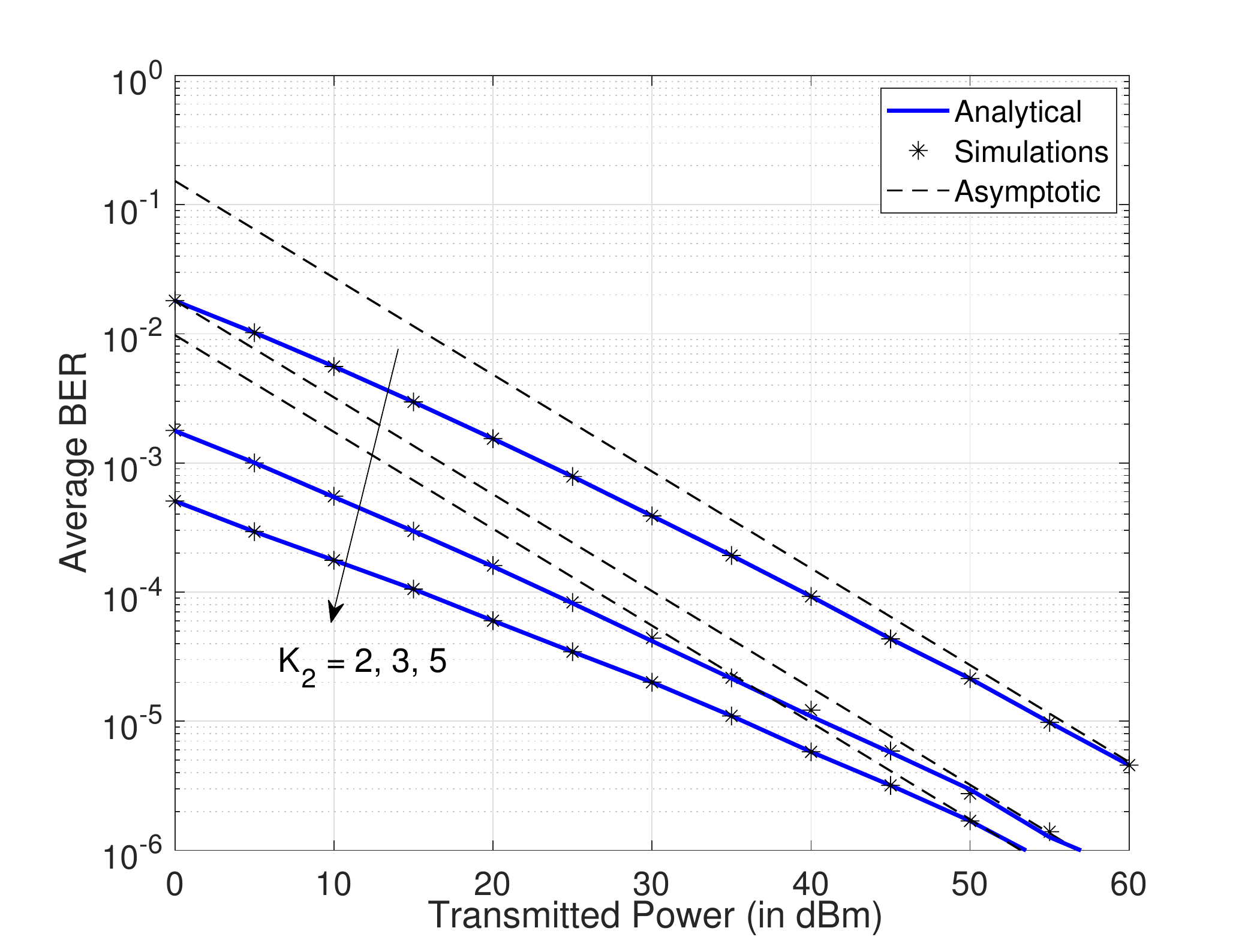}} \hspace{-6mm}
	\caption{Performance demonstration of multi-RIS transmissions for R2V link.}
	\label{fig:rf_outage}
	\label{fig:rf_ber}
\end{figure*}

\begin{figure*}[!htbp]
	\centering
	\subfigure[Outage probability.]{\includegraphics[scale=0.35]{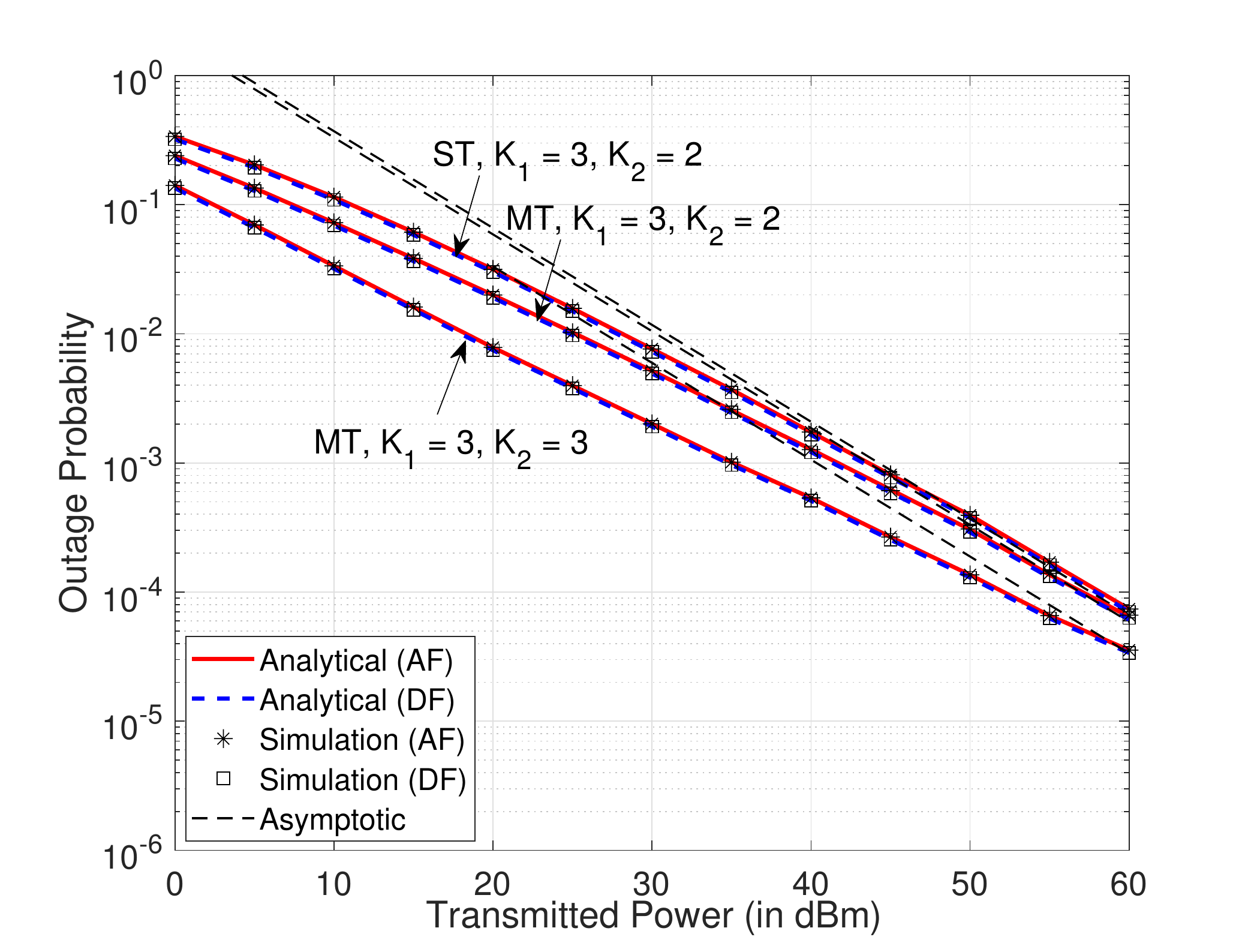}} \hspace{-6mm}
	\subfigure[Average BER. ]{\includegraphics[scale=0.35]{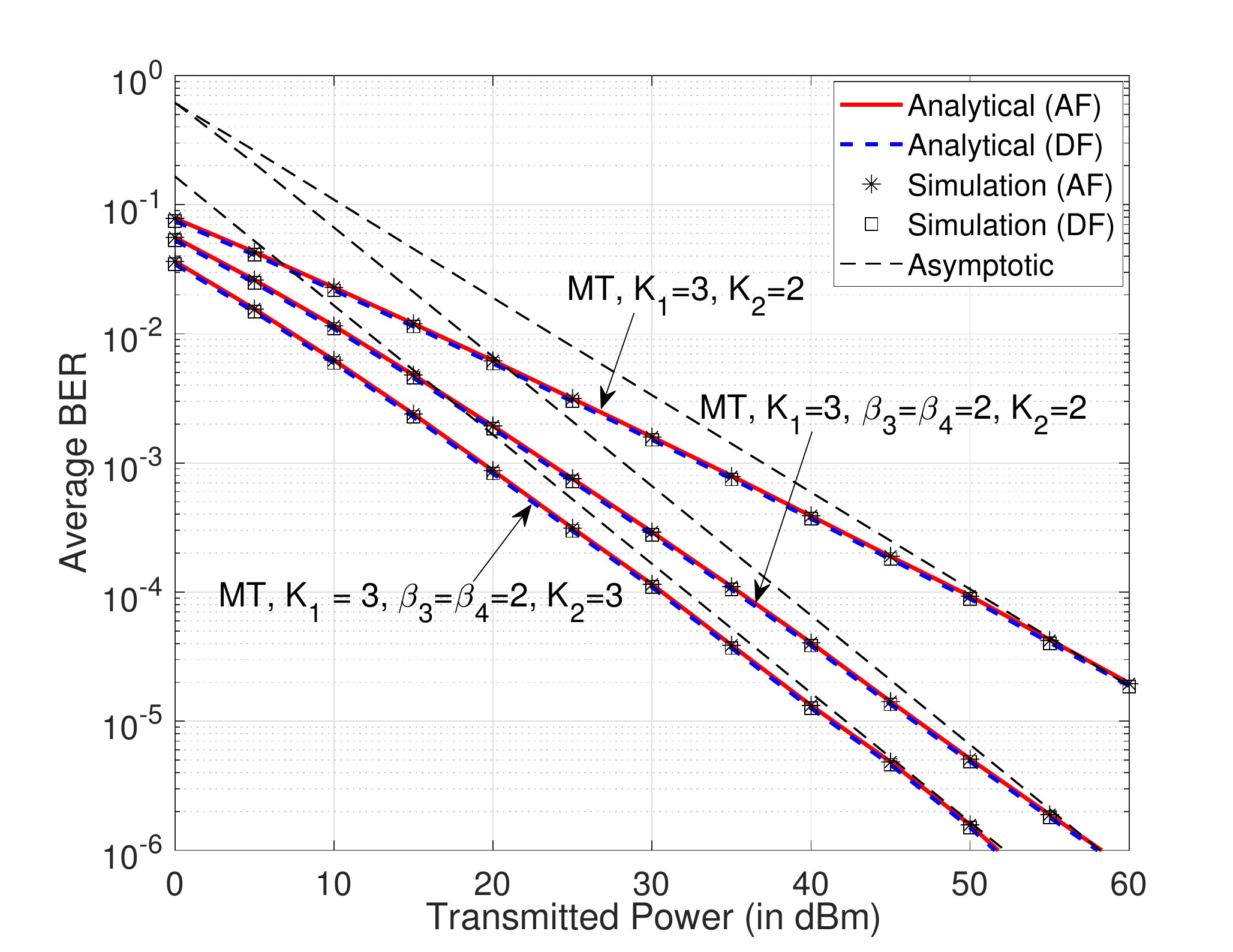}} \hspace{-6mm}
	\caption{Performance demonstration of RIS-assisted multi-hop mixed FSO-R2V system.}
	\label{fig:relay_outage}
\end{figure*}

\section{Numerical and Simulation Analysis}\label{sec:sim_results}
\begin{figure}[!htbp]
	\centering
	{\includegraphics[scale=0.4]{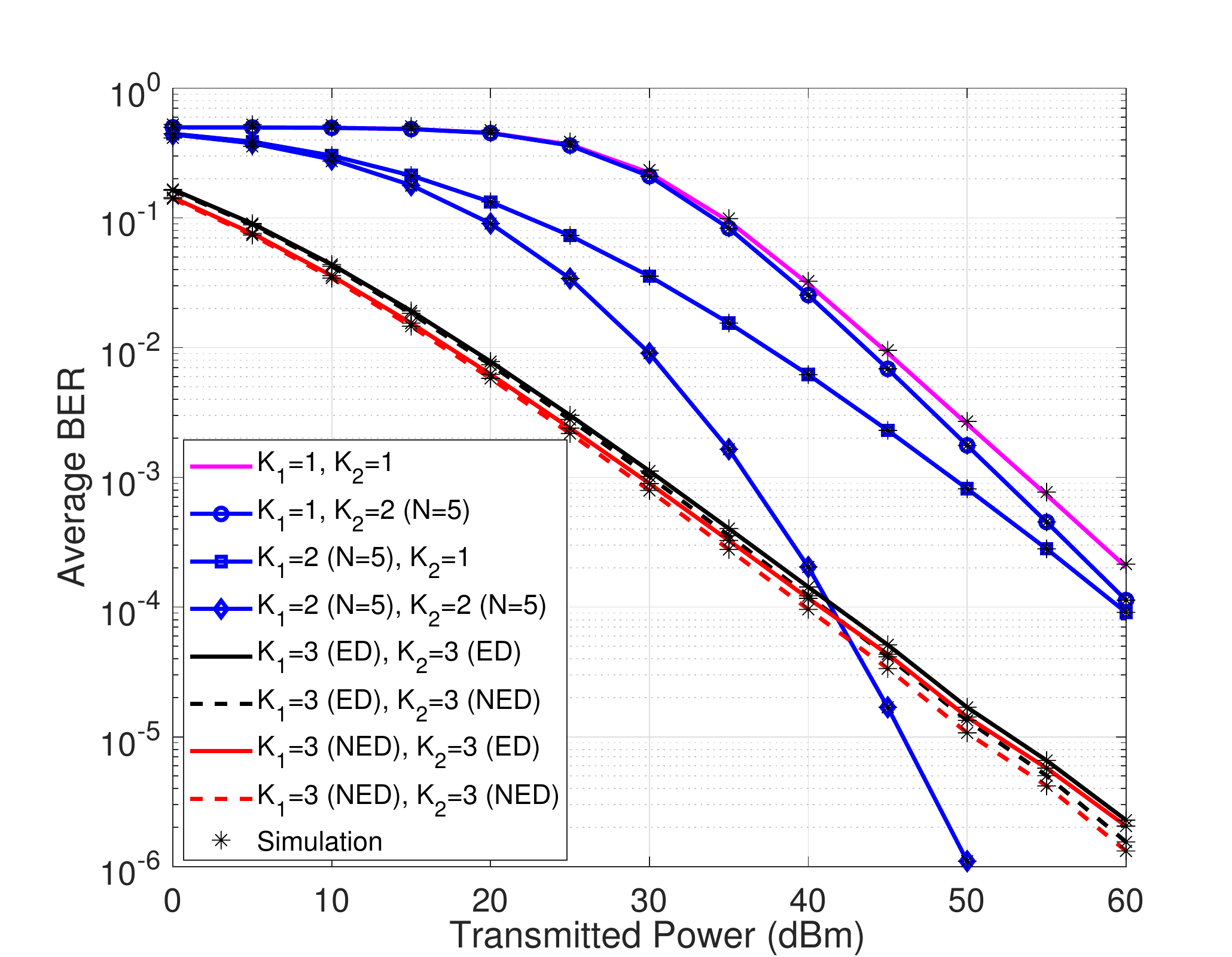}} \hspace{-6mm}
	\caption{Performance comparison of RIS-assisted multi-hop mixed FSO-R2V system with other system configurations. Here, $\{K_1=1, K_2=1\}$: without FSO-RIS and without RF-RIS ; $\{K_1=1, K_2=2 (N=5)\}$: without FSO-RIS and single RF-RIS with $5$ elements; $\{K_1=2 (N=5), K_2=1\}$: single FSO-RIS with $5$ elements and  without RF-RIS; $\{K_1=2 (N=5), K_2=2 (N=5)\}$: single FSO-RIS with $5$ elements and single RF-RIS with $5$ elements; $K_1=3$ and $K_2=3$:  proposed multihop (i.e. $2$ FSO-RIS with single element and $2$ RF-RIS with single element);  ED: equidistant placement of RISs; NED:  non-equidistant placement of  RISs; $d_1=1000$\mbox{m} and $d_2= 300$\mbox{m}. NED for FSO has first RIS at $200$\mbox{m} and second RIS at $500$\mbox{m}, and NED for RF has first RIS at $40$\mbox{m} and second RIS at $160$\mbox{m}.}
	\label{fig:diff_distance_ber}
\end{figure}

This section demonstrates the performance of multiple RIS empowered vehicular communications and  validates the mathematical performance metrics using  Monte-Carlo simulations. We use MATLAB and MATHEMATICA computational software for computer simulations. We also demonstrate the tightness of asymptotic expressions at a high SNR and verify the diversity order of the system by varying channel fading parameters.  We assume that the backhaul FSO link length is $d_{1}=1000$ \mbox{m} (a typical backhaul scenario for FSO in urban area) and $K_1-1$ optical RISs are deployed at equal distances. We also show the effect of unequal deployment of RIS-FSO in Fig.~\ref{fig:diff_distance_ber}. We model the FSO atmospheric  turbulence using  dGG parameters corresponding to strong turbulence (ST) ($\alpha_{1}=1.8621$,$\alpha_{2}=1$,$\beta_{1}=0.5$,$\beta_{2}=1.8$,$\Omega_{1}=1.5074$,$\Omega_{2}=0.928$) and moderate turbulence (MT) ($\alpha_{1}=2.169$, $\alpha_{2}=1$, $\beta_{1}=0.55$, $\beta_{2}=2.35$, $\Omega_{1}=1.5793$, $\Omega_{2}=0.9671$)\cite{AlQuwaiee2015}.   We compute the effective path loss for FSO links for  a visibility range of $3$ \mbox{Km} \cite{Kim2001} (a commonly used value in hazy conditions) at a wavelength $1550$ \mbox{nm}. We take values of  detector responsitivity is $0.41$ \mbox{A/W} and AWGN noise variance of  $10^{-14}$ \mbox{$A^{2}$/GHz}. We consider pointing error parameter $\rho=2.5$ at $K_{1}=2$ and low or negligible pointing error ($\rho=5$) at higher $K_{1}$ due to perfect beam alignment.

To illustrate the vehicular communications, we use the   link distance of $d_{2}=100$\mbox{m} with the orientation of   RF-RIS, as shown in 	Fig.~\ref{fig:system_model} by deploying $K_2-1$ RISs at  equal distances. The use of $100$\mbox{m} RIS-assisted vehicular link is considered to accommodate a typical cluttered environment in high-traffic scenarios, for example, near the shopping mall and traffic light junctions. We use the equal distance placement of RF-RIS for the proposed scheme to have a higher resultant path loss \cite[eq. 5]{Tang_2021} for better performance comparison.  In Fig.~\ref{fig:diff_distance_ber}, we also considered a higher vehicular link of $300$ \mbox{m} and  compared the performance between unequal and  equal distance placement of RF-RIS modules.    We use RF dGG parameters $\alpha_{3}=1.5$,$\alpha_{4}=1$,$\beta_{3}=1.5$,$\beta_{4}=1.5$,$\Omega_{3}=1.5793$,$\Omega_{4}=0.9671$ to model the R2V link for  vehicular transmissions \cite{Petros2018}. We use shadowing parameter $m=1.2, 7.4, 15$ corresponding to $K_{2}=2, 3, 5$ hops to model the the effect of shadowing in the presence of obstacles \cite{Rami2021IGG}. The lower the parameter $m$, the higher the effect of shadowing. In addition, we also vary the path loss exponent $a$ w.r.t $K_{2}$ and in particular, we consider $a=4, 3, 2$ at $K_{2}=2, 3, 5$ respectively. For the direct link, we use same fading parameters with shadowing $m=1.2$ and path loss exponent $a=4$. To simulate the path gain of the R2V link, we  use equal gain of  transmit and receive antenna  as  $G_{t}=G_{r}=25$ \mbox{dBi} at a  frequency of $800$ \mbox{MHz}. These parameters are selected to ensure that the both FSO and R2V have similar average SNR to avoid imbalance of the two links.  Indeed, the change in carrier frequency impacts the path loss, and a different carrier frequency will change the results without a significant change in the comparative study.  The variance of AWGN is considered to be  $-104.4$ \mbox{dBm} with a  channel bandwidth of  $20$ \mbox{MHz}. For RF-RIS, the wavelength at 800 MHz is $37.5$ cm. Since the dimension of RIS element is  proportional to wavelength, the RIS element has a dimension of  $37.5 \mbox{cm} \times 37. 5 \mbox{cm}$. The size of FSO RIS is equal to the  diameter of the receiver aperture, usually taken as   $20\mbox{cm}$  \cite{Najafi2019}.

First, we demonstrate the effect of multiple RISs  for both FSO (with moderate turbulence) and vehicular transmissions by plotting outage probability (in Fig.~\ref{fig:outage}(a)), and  average BER for differential binary phase shift keying ($p=1$ and $q=1$)(in Fig.~\ref{fig:ber}(b)) considering the same turbulence,  pointing errors, and fading scenarios in each hop. Figures show the performance  of multi-RIS  system versus number of RISs ($K_{1}=K_{2}=K$, $K-1$ RISs constituting $K$ hops) for FSO and R2V transmissions at a transmit power of $P_{t}=30$ \mbox{dBm} and $P_{t}=50$ \mbox{dBm}.  We consider two pointing error (PE) scenarios: PE1, and PE2. In PE1, we assume that  all the FSO links undergo the same pointing errors  $\rho=2.5$. In PE2, we use higher pointing error parameters ($\rho=1$  or $\rho=2.5$)  from the source transmitter to the first optical RIS and last optical RIS to the relay  and negligible   pointing errors  involving RIS to RIS due to a perfect beam alignment. The figure shows that an increase in the number of RIS modules between a source and destination decreases the performance, which can be regarded as a counter productive. However, without the use of multiple RIS, there exists no direct link which precludes any FSO transmissions and highly degraded R2V links. Thus, the use of multi-RIS should be limited enabling line-of-sight transmissions. Note that the reason for degradation in the performance with an increase in the number of RIS  modules is the manifestation of cascading of channel coefficients comprising  deterministic path gain ($h_l<1$). It should be mentioned that the use of multiple RIS is applied in the same channel conditions without harnessing the LOS, depicting a degradation in the performance with multiple RIS. For a fair comparison in the multihop scenario, an increase in the number of RIS may enjoy better channel conditions resulting in improved performance, as illustrated below.

Next, we consider a more practical setup with different fading scenarios in each hop to demonstrate the use of multiple RISs for FSO  (Fig.~\ref{fig:fso_outage}(a) and Fig.~\ref{fig:fso_outage}(b)) and RF transmissions (Fig.~\ref{fig:rf_outage}(a) and Fig.~\ref{fig:rf_outage}(b)). We plot the outage performance of FSO system for ST scenario and for different hops $K_{1}=2, 3, 4$ in Fig.~\ref{fig:fso_outage}(a). It can be seen that an increase in the number of optical RIS modules between the source and relay increases the performance due to a decrease in the effect of pointing errors with close placement of RIS modules. In particular, with $K_{1}=2$, FSO links suffer from higher pointing errors $\rho=2.5$ but when the number of optical RIS modules are increased to $2$ for $K_{1}=3$, FSO signal in each hop suffers from relatively lesser pointing error $\rho=5$. Further, with $K_{1}=4$ i.e., $3$-optical RIS modules between source and relay, FSO transmissions have negligible or no pointing errors in each hop. Thus the multi-RIS deployment improves the FSO system performance with an increase in the number of hops or optical RIS modules. Moreover, lesser transmit power is needed with an increase in the optical RIS modules to achieve a desired performance. For example, for an outage performance of $10^{-3}$, we need $30$\mbox{dBm} of transmit power with $K_{1}=4$ but $P_{t}=35$\mbox{dBm} and $P_{t}=40$\mbox{dBm} for $K_{1}=3$ and $K_{1}=2$ hops respectively. The figure also depicts that the diversity order or the slope of the outage plots is the same for different $K_{1}$-hops as the diversity order  is determined by the atmospheric turbulence parameters (ST) independent of pointing errors.

The average BER performance of multi-hop optical RIS FSO system is shown in Fig.~\ref{fig:fso_outage}(b) for ST and MT atmospheric turbulence scenarios with $K_{1}=3, 4$. Here, BER performance improves with an increase in the number of optical RIS modules due to negligible or no pointing error at higher $K_{1}$. It can be seen that strong turbulence requires an additional  $15$ \mbox{dBm}  of the transmit power  compared with the moderate scenario to achieve an outage probability of $10^{-4}$ for $K_{1}=3$ and $20$ \mbox{dBm} for $K_{1}=4$. The plot also shows that the diversity depends on FSO system parameters as the average BER plots are steeper for MT (diversity order $G_{\rm BER}=1.12$) when compared with ST scenario ($G_{\rm BER}=0.93$).

Under the similar practical setup, we illustrate the outage probability Fig.~\ref{fig:rf_outage}(a) and  the average BER performance  
Fig.~\ref{fig:rf_outage}(b) for multi-hop RF system. Here, we consider higher shadowing and also larger path loss exponent $a$ for lower $K_{2}$ and subsequently comparatively  lesser parameters for higher $K_{2}$ to depict the scenario of good channel conditions with an increase in  RIS modules. The figure depicts that  there is an enhancement in the RF system performance  with an increase in the number of RIS modules or hops. For a comparison, we have also plotted multi-hop DF relay performance in Fig.~\ref{fig:rf_outage}(a) using  path loss exponent $a=4$ for all $K_{2}$. It can be seen that the DF relay performance is better than that of multi-hop RIS system requiring more energy consumption compared with the RIS-based system. We have also plotted a single RIS system with 2-elements (assuming perfect phase compensation) performing better at a high SNR. Note that  we consider a single-element RIS since  it is hard for phase compensation  for multiple-element RIS in the multi-hop scenario. In Fig.~\ref{fig:rf_outage}(b), we have shown the average BER performance for different $K_{2}$. Here, the average BER performance improves with an increase in $K_{2}$ due to reduced effect of shadowing and path loss. As $K_{2}$ increases, we require a less transmit power  to achieve the desired performance. For example, a transmit power of $10$\mbox{dBm} is needed at $K_{2}=3$ when compared with $25$\mbox{dBm} at $K_{2}=2$ to achieve a desired  average BER performance of $10^{-3}$.

In Fig.~\ref{fig:relay_outage}(a) and \ref{fig:relay_outage}(b),  we demonstrate the impact of multiple RIS modules on the performance of a  mixed FSO-RF system. We consider both ST and MT with $K_{1}=3$ hops in the FSO links. We  plot the  outage probability as depicted in  Fig.~\ref{fig:relay_outage} (a) for both DF and AF relay  by considering $K_{1}=3$ hops in the FSO link and $K_{2}=2$, $K_{2}=3$   RF links to demonstrate the improvement in performance with an increase in $K_{2}$. Finally in Fig.~\ref{fig:relay_outage} (b), we simulate the average BER performance of mixed FSO-RF system with MT, $K_{1}=3$ in FSO links and $K_{2}=2, 3$ in RF links. We also vary the parameter $\beta$ to depict that the diversity order of the system depending on fading parameters. It can be seen from the figure  that an increase in the parameter $\beta$ improves the performance due to the decrease in the fading severity.

Finally, we compare the average BER performance of our proposed system with other system configurations, as described in the caption of Fig.~\ref{fig:diff_distance_ber}, for both equal and unequal placement of RIS modules. The results demonstrate that the placement of RIS modules in the equidistant (ED) and non-equidistant (NED) configurations (represented by the black and red plots, respectively) have negligible impact on the average BER performance of our multihop scheme. However, the unequal placement of RIS modules performs slightly better than the equidistant scenario due to lower path loss. This finding may motivate further investigation into optimizing the placement distance to enhance performance further. We also demonstrate the average BER performance of single RIS modules containing five elements (illustrated by three blue-colored plots). The results show that incorporating RIS modules in FSO and RF links yields better performance than employing RIS solely in FSO or RF. The illustration indicates that the suggested multihop transmission approach (employing 2 RIS modules with one element each in FSO and RF connections) outperforms the single RIS module method (featuring five elements each in FSO and RF connections) for a usable transmission power of up to $40$ \mbox{dBm}. Moreover, using RIS modules leads to improved outcomes compared to situations without RIS implementation, as demonstrated by the magenta plots.

\section{Conclusions}\label{sec:conc}
We investigated the  multiple  RIS empowered  mixed FSO-RF multihop system for vehicular communications. We developed a  framework to derive PDF and CDF of cascaded channels for generalized fading models by considering single-element RIS avoiding complex procedure for phase compensation compared with the multiple-element RIS system under multi-hop scenario. We employed  AF and DF relaying protocols to mix two different technologies and presented  system performance  by deriving analytical expressions for the outage probability and average BER. We also analyzed asymptotic behavior of the outage probability in the high SNR  using Gamma functions to reveal insight on the effect of  channel parameters on the system performance. Simulation results  demonstrated the effectiveness of multiple RISs modules  to  achieve  LOS propagation for FSO transmissions and reliable connectivity  for  vehicular communication. In FSO links, the use of multiple RIS modules reduces the pointing error in each hop due to perfect beam alignment with subsequent use of multiple RIS. With the multiple RF-RIS deployment, the effect of deep shadowing can be reduced due to close proximity of RISs.  	
	The suggested multi-hop transmission approach may surpass the single RIS module technique in specific situations of practical importance. The suggested multiple RIS based communications can be a suitable candidate to provide  seamless connectivity for autonomous vehicular systems. 
Although the relaying system performance is better, the RIS-based system is more energy efficient and requires less power consumption.

In the future, the proposed work could encompass a thorough, measurement-based channel analysis for multihop-based RIS systems to improve performance evaluation. Additionally, optimizing the positioning of RIS modules may present a feasible solution for boosting overall performance.

\section*{Appendix A}
Most of the works in the literature use induction based approach to compute the PDF of product of random variables. However, similar analysis cannot be used for all the generalized fading models. Thus, we employ Mellin's transform to compute the PDF of the product of random variables. The Mellin's transform of a function $f_X(x)$ is  given as \cite{Kilbas_FoxH}
\begin{equation}\label{eq:gen_mellin}
M(r) = M\{f_X(x)\} = \int_{0}^{\infty} f_X(x) x^{-r} \diff x 
\end{equation}
The inverse Mellin's transform as
\begin{equation}\label{eq:gen_inv_mellin}
f_X(x) = \frac{1}{x} \frac{1}{2\pi \J} \int_{\mathcal{L}} M(r) x^{-r} \diff r 
\end{equation}
If $f_X(x)$ is the PDF then its Mellin's transform is nothing but the $r^{th}$-moment $\mathbb{E}[X_{}^{r}]$.	Thus, we use the Mellin's  transform to find the PDF of the product of $K$ random variables  as \cite{Kilbas_FoxH}
\begin{equation}\label{eq:gen_prod_pdf_1_mellin}
f_X(x) = \frac{1}{x} \frac{1}{2\pi \J} \int_{\mathcal{L}} \prod_{i=1}^{K} \Aver{X_{i}^{r}} x^{-r} \diff r 
\end{equation}
Substituting \eqref{eq:gen_pdf} in  $\Aver{X_{i}^{n}} = \int_{0}^{\infty} x^{r} f_{X_{i}}(x) dx$ and using the identity \cite[{eq.} 2.8]{M-Foxh}, the $r$-th moment can be computed as
\begin{align}\label{eq:moment_gen_prod_pdf_1}
&	\Aver{X_{i}^{r}} = \psi_{i} \int_{0}^{\infty} x^{r+\phi_{i}-1} H_{p,q}^{m,n} \bigg[\begin{array}{c}\zeta_{i} x \end{array} \big\vert \begin{array}{c}\{(a_{i,j},A_{i,j})\}_{j=1}^{p}\\ \{(b_{i,j},B_{i,j})\}_{j=1}^{q} \end{array}\bigg] dx \nonumber \\&
\hspace{0mm} = \psi_{i} \zeta_{i}^{-r-\phi_{i}} \frac{\prod_{j=1}^{m}\Gamma(b_{i,j}+B_{i,j}(r+\phi_{i}))}{\prod_{j=n+1}^{p}\hspace{-2mm}\Gamma(a_{i,j}+A_{i,j}(r+\phi_{i}))} \nonumber \\ &\frac{\prod_{j=1}^{n}\Gamma(1-a_{i,j}-A_{i,j}(r+\phi_{i}))}{\prod_{j=m+1}^{q}\hspace{-2mm}\Gamma(1-b_{i,j}-B_{i,j}(r+\phi_{i}))}
\end{align}
\normalsize

\normalsize
We use \eqref{eq:moment_gen_prod_pdf_1} in \eqref{eq:gen_prod_pdf_1_mellin} and apply the integral representation of   the Fox's-H function to get \eqref{eq:gen_prod_pdf}.
Using \eqref{eq:gen_prod_pdf} in $F_{X}(x) = \int_{0}^{x}f_{X}(t) dt$, an expression for the CDF:
\begin{align}\label{eq:gen_prod_cdf_1}
&	F_{X}(x) =\prod_{i=1}^{K} \psi_{i} \zeta_{i}^{-\phi_{i}}  \frac{1}{2\pi \J} \int_{\mathcal{L}} (\prod_{i=1}^{K}\zeta_{i})^{r} (\int_{0}^{x} t^{-1+r} dt)\nonumber \\& \prod_{i=1}^{K} \frac{\prod_{j=1}^{m}\Gamma(b_{i,j}+B_{i,j}(-r+\phi_{i}))}{\prod_{j=n+1}^{p}\Gamma(a_{i,j}+A_{i,j}(-r+\phi_{i}))}\nonumber \\& \frac{\prod_{j=1}^{n}\Gamma(1-a_{i,j}-A_{i,j}(-r+\phi_{i}))}{\prod_{j=m+1}^{q}\Gamma(1-b_{i,j}-B_{i,j}(-r+\phi_{i}))}   \diff r
\end{align}

Using the inner integral  $\int_{0}^{x} t^{-1+r} dt = \frac{x^{r}}{r} = x^{r} \frac{\Gamma(r)}{\Gamma(1+r)}$ in in \eqref{eq:gen_prod_cdf_1}, and applying the definition of Fox's-H function, we get \eqref{eq:gen_prod_cdf}, which concludes the proof.

\bibliographystyle{IEEEtran}
\bibliography{Multi_RISE}

\end{document}